\documentclass[floatfix,flushbottom,nobibnotes,nofootinbib,amsmath]{PoS}
\usepackage{color}
\usepackage{subfigure}
\usepackage{multicol}
\usepackage{epsfig}
\usepackage{wrapfig}
\usepackage{lipsum}
\usepackage{graphicx}
\usepackage{media9}
\usepackage{animate}
\usepackage{bm}
\usepackage[export]{adjustbox}

\newcommand\bef{\begin{figure}}
\newcommand\eef[1]{\label{fg:#1}\end{figure}}
\newcommand\bec{\begin{center}}
\newcommand\eec{\end{center}}
\newcommand\besf{\begin{subfigure}}
\newcommand\eesf[1]{\label{sfg:#1}\end{subfigure}}
\newcommand\beq{\begin{equation}}
\newcommand\eeq[1]{\label{#1}\end{equation}}
\newcommand\beqa{\begin{eqnarray}}
\newcommand\eeqa[1]{\label{#1}\end{eqnarray}}
\newcommand\bet{\begin{table}}
\newcommand\eet[1]{\label{tb:#1}\end{table}}
\newcommand\best{\begin{subtable}}
\newcommand\eest[1]{\label{stb:#1}\end{subtable}}
\newcommand\betb{\begin{center}\begin{tabular}}
\newcommand\eetb{\end{tabular}\end{center}}
\newcommand\beit{\begin{itemize}}
\newcommand\eeit{\end{itemize}}

\newcommand\incfig[2]{\includegraphics[scale=#1]{./#2}}

\definecolor{DarkGreen}{rgb}{0.00,0.29,0.00}
\definecolor{DarkRed}{rgb}{0.79,0.00,0.00}

\def\prsp#1#2%
  {\mathop{}%
   \mathopen{\vphantom{#2}}^{#1}%
   \kern-\scriptspace%
   #2}

\title{Hadron Spectroscopy and Resonances: Review}

\ShortTitle{Hadron Spectroscopy and Resonances: Review}

\author{\speaker{M. Padmanath}\\
        Instit\"ut f\"ur Theoretische Physik, Universit\"at Regensburg, 93053 Regensburg, Germany\\
        E-mail: \email{Padmanath.M@physik.uni-regensburg.de}}

\abstract{I review recent results on hadron spectroscopy using lattice QCD. In light of 
the discoveries in heavy baryon sector at LHCb over the past few years, lattice calculations 
in this regard are emphasized. Investigations on light baryon, heavy-heavy and heavy-light 
meson resonances are also discussed.}

\FullConference{The 36th Annual International Symposium on Lattice Field Theory - LATTICE2018\\
		22-28 July, 2018\\
		Michigan State University, East Lansing, Michigan, USA.}

\begin{document}

\section{Introduction}\label{Intro}

The observation of many interesting resonance structures in hadronic final states in recent years 
has spurred huge interest in hadron spectroscopy investigations both theoretically and experimentally. 
The discovery of a doubly charmed baryon $\Xi^{+}_{cc} (ccu)$ with a mass of $3621.40\pm0.78$ MeV 
by the LHCb Collaboration \cite{Aaij:2017ueg} marks an important milestone in our quest to understand Quantum 
ChromoDynamics (QCD). Another highlight is the unambiguous observation by the LHCb Collaboration of five 
new narrow $\Omega_c$ resonances in $\Xi^{+}_{c}K^{-}$ invariant mass distribution in the energy range 
between $3000-3120$ MeV \cite{Aaij:2017nav}. Four out of these five resonances have been later confirmed 
by the Belle Collaboration \cite{Yelton:2017qxg}. Compelling evidence for charged meson-like resonances 
in the heavy quarkonium energy range indicate the existence of four-quark bound states (for review see 
Refs. \cite{Lebed:2016hpi,Esposito:2016noz,Olsen:2017bmm}). Similarly, the LHCb discovery of exotic 
structures in the $J/\psi~p$ channel confirms the existence of charmonium-nucleon pentaquark resonances 
\cite{Aaij:2015tga}. Measuring the properties of these resonances can further enhance our understanding 
of QCD. Experiments are engaged in the search for new resonances with an aim of precisely measuring their 
mass, lifetime, production and decay mechanisms, etc. \cite{Yelton:2018mag,Aaij:2018gfl,Aaij:2018wzf}. 
This motivates theoretical investigations to make predictions for 
such resonances, that can be put to test in various experiments such as LHCb, Belle.

Lattice QCD calculations of hadron spectroscopy have achieved remarkable progress over the past ten years 
in making large volume simulations with physical quark masses, impressive statistical precision and good 
control over different systematic uncertainties \cite{Aubin:2004wf,Durr:2008zz,Bazavov:2009bb,Dowdall:2012ab,
Namekawa:2013vu,Borsanyi:2014jba}. Several lattice groups have been performing detailed 
systematic investigations of stable hadrons that are well below the lowest allowed strong decay threshold. 
A discussion on various lattice systematic uncertainties and how they are addressed by different lattice 
groups is made in Ref. \cite{Liu:2016kbb}. Multiple exploratory studies are also being performed in order 
to understand and to help interpreting these observed hadrons close to strong decay thresholds and resonances 
above the threshold. A detailed review on methodologies for treating the hadronic resonances on the lattice 
and various lattice calculations along these lines can be found in Ref. \cite{Briceno:2017max}.

In this report, I review the recent results from lattice calculations with emphasis on those that are relevant 
considering the present and future experimental progress. Section \ref{methodology} briefly outlines the basic 
lattice methodology that is relevant for the results presented in this review. In Section \ref{ground}, the high 
precision lattice results for masses of stable hadrons are summarized. The studies of excited hadrons, that 
have been in the scientific limelight considering recent experimental discoveries, are discussed in 
Section \ref{excited}. Sections \ref{mesres} and \ref{barres} address recent lattice calculations of hadronic 
resonances involving rigorous finite volume analysis. The recent lattice results on unconventional hadrons 
that do not fit into $\bar qq$ and $qqq$ picture are presented in Section \ref{beyond}. Section \ref{summar} 
summarizes the review. 


\section{Lattice methodology}\label{methodology}
The physics of hadrons is commonly extracted on the lattice from the finite volume Euclidean correlation functions.
In order to study the hadron spectrum, one computes the two point correlation functions,
\begin{equation}
C_{ij}(t_f-t')=\langle O_{i}(t_f)O^{\dagger}_{j}(t') \rangle=\sum_{n} \frac{Z_i^{n}Z_j^{n*}}{2 E_n}e^{-E_n(t_f-t')},
  \label{eq:2-1}
\end{equation}
between hadronic currents ($O_{i}(t)$) that are built to respect the quantum numbers of interest. 
The operator $O_{i}(t)$ can couple to all the states, including single-particle levels as well as 
multi-particle levels and their radial excitations, with these quantum numbers. A general practice to 
extract the excited spectrum is to compute matrices of correlation functions between a basis of 
interpolators $O_{i}(t)$ \cite{Basak:2005ir,Basak:2005aq,Dudek:2010wm,Thomas:2011rh} and to solve the 
Generalized EigenValue Problem (GEVP)~\cite{Michael:1985ne,Luscher:1985dn,Blossier:2009kd}
\begin{equation}
C_{ij}(t)v_j^n(t-t_0)=\lambda^n(t-t_0)C_{ij}(t_0)v_j^n(t-t_0).
  \label{eq:2-2}
\end{equation}
Energies ($E_n$) are extracted from exponential fits to the large time behavior of the eigenvalues 
$\lambda^n(t-t_0)$. The operator state overlaps $Z_i^{n} = \langle O_{i}|n\rangle$ are related to 
the eigenvectors $v_j^n(t-t_0)$. 

Hadron masses that are well below the lowest allowed strong decay threshold and hence deeply bound are 
trivially related to the lattice energies $m_H=E_{lat}({\bf p}={\bf 0})$ up to exponentially suppressed 
corrections. Lattice investigations of these stable hadrons require only the extraction of ground states. 
Simple fits to the large time behavior of the correlation functions provide quite precise estimates for masses 
of these hadrons. This has been in practice, since the early applications of lattice QCD in hadron 
spectroscopy. There is a number of ground state baryons that are stable to strong decays and can be studied 
quite precisely on the lattice. 

Most hadrons appear above or close to the strong decay threshold.  There is no direct procedure to extract 
the energies for these hadrons from the discrete spectrum on the lattice \cite{Maiani:1990ca}. Properties 
of these near or above threshold hadron excitations have to be inferred from the infinite volume scattering 
matrices. L\"uscher's finite volume method is a widely used approach to extract the infinite volume 
scattering matrix from the discrete finite volume energy spectrum \cite{Luscher:1986pf,Luscher:1990ux,Luscher:1991cf}. 
This approach relates the infinite volume phase shifts, that possess all information on the scattering process, 
to the discrete energy spectrum in the finite volume through known kinematic functions. Several extensions and 
generalizations to L\"uscher's original proposal have been made over the past five years: {\it e.g.} to 
describe the two particle scattering with different particle identities, in different boundary conditions, 
in moving frames, multiple partial waves, coupled channel scattering and more (see Refs. \cite{Briceno:2014oea,Briceno:2018bnl}). 
Updates on efforts to build extensions that relate discrete finite volume energy spectrum to three-body 
scattering amplitudes have been reported in this conference \cite{Blanton:2018guq,Mai:2018xwa}. A detailed 
discussion on the formalism, its extensions and a comprehensive list of references can be found in Ref. \cite{Briceno:2017max}.

There exist other formalisms to extract infinite volume scattering information from the lattice such as the 
HALQCD method \cite{Ishii:2006ec,HALQCD:2012aa}, finite volume Hamiltonian EFT \cite{Hall:2012pk} and the 
relatively new optical potential method \cite{Agadjanov:2016mao}. Attempts to relate the three-body scattering 
amplitudes with the discrete energy spectrum in the finite volume are also being made \cite{Hammer:2017uqm,
Hammer:2017kms,Doring:2018xxx,Mai:2018djl}. For detailed information on these formulations the reader may 
refer to the original articles, while a summary of lattice investigations following these approaches can be 
found in the reviews from previous lattice conferences \cite{Liu:2016kbb,Prelovsek:2014zga}.

Most of the results using a finite volume analysis presented in this review are based on L\"uscher's 
formalism and its extensions. Novel lattice QCD techniques such as distillation \cite{Peardon:2009gh}, 
stochastic distillation \cite{Morningstar:2011ka} and unbiased noise reduction techniques \cite{Bali:2009hu} 
to compute the all-to-all quark propagation diagrams have made lattice calculations using up to two hadron 
interpolators and involving rigorous finite volume analysis possible. Lattice studies of resonances 
discussed in this review are performed using such techniques and consider all required Wick contractions, 
except those with OZI suppressed heavy quark self-annihilation diagrams, where applicable, to compute 
the correlation matrices. There is, as yet, no numerical study of hadronic resonances using three hadron 
interpolators, except for investigations of nuclear binding energies. 


\section{Ground state hadrons}\label{ground}

Postdicting the mass of ground state hadrons is a standard benchmark procedure that boosts confidence 
in our methodology and hence lends support to predictions. Several calculations have been performed with 
good control over the statistical as well as systematic uncertainties and have precisely predicted/postdicted 
meson ground state masses ({\it c.f.} Ref. \cite{Dowdall:2012ab}). Most of these have also been discussed 
in previous lattice conferences \cite{Prelovsek:2014zga}. In what follows, the emphasis is given on the 
recent high precision lattice results on the light as well as heavy ground state baryons.     

\begin{figure}
\includegraphics[height=4.5cm,width=7.2cm]{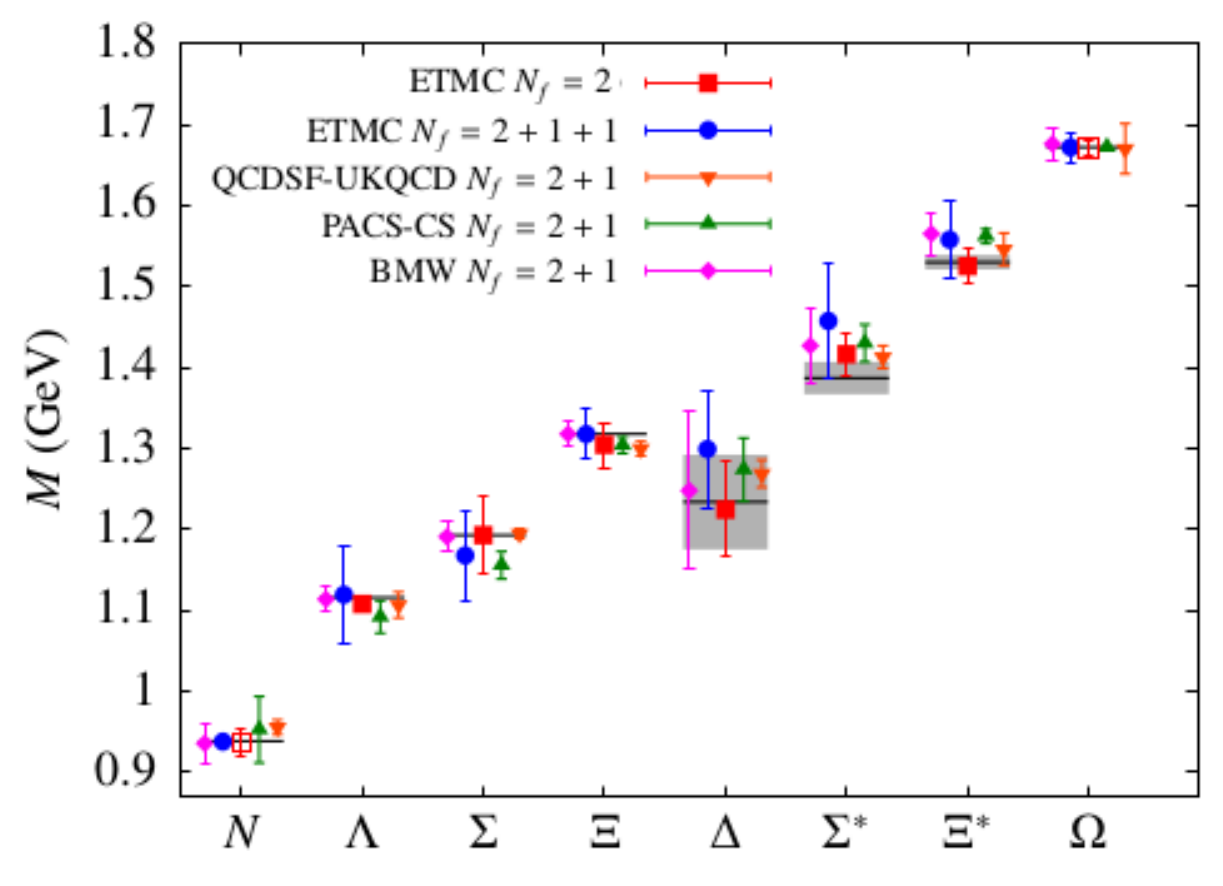}
\includegraphics[height=4.8cm,width=7.6cm]{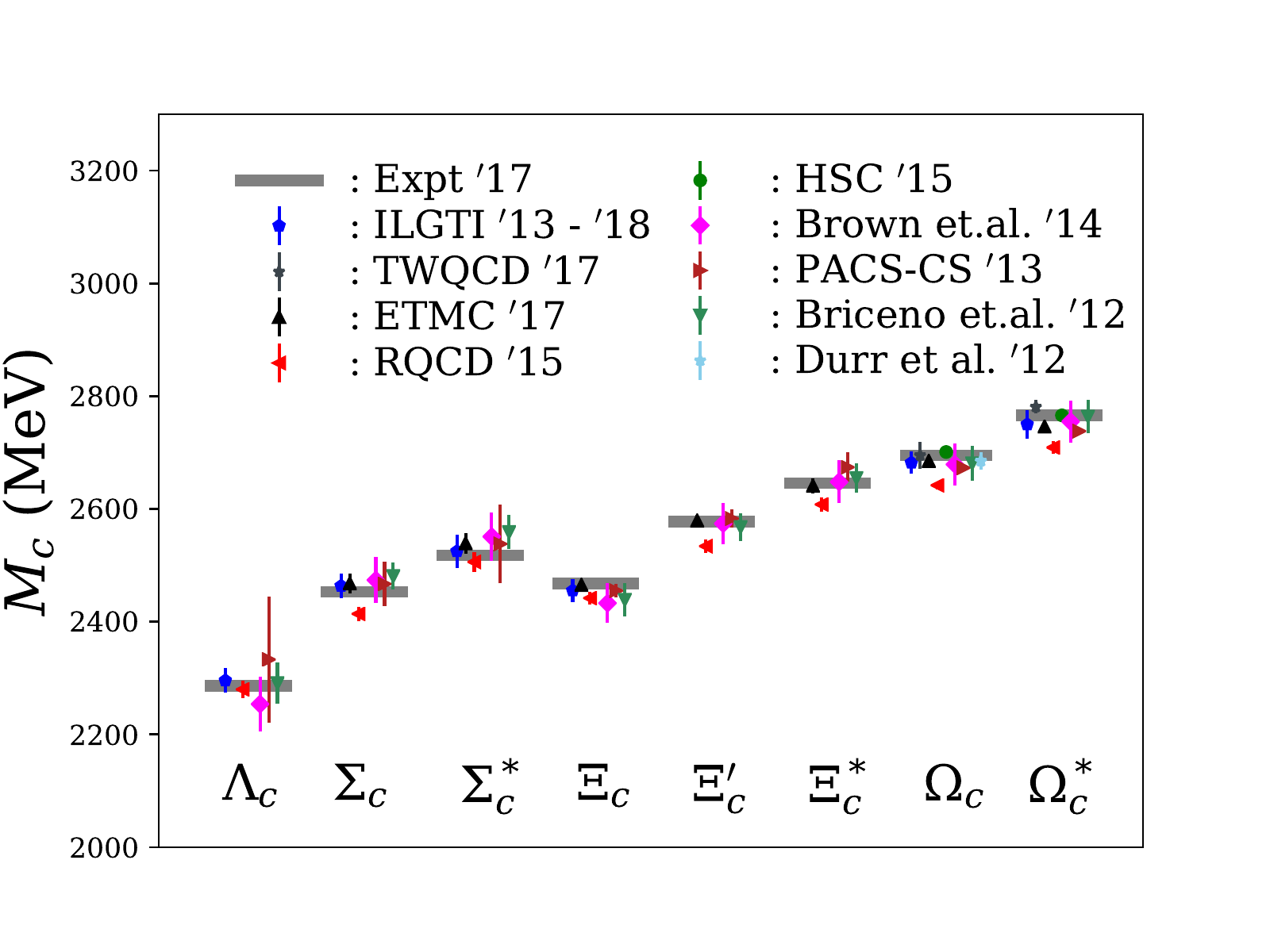}
\caption{Left: (figure adapted from Ref. \cite{Alexandrou:2017xwd}) Lattice results for the octet and 
decuplet light baryon masses. The horizontal lines represent the experimental masses and the bands 
represent the widths. Unfilled symbols indicate baryon masses that have been used as inputs to the 
calculation, whereas the filled symbols refer to postdictions. Right: Lattice results for single charmed 
baryon masses.}\label{lcbaryons}
\end{figure}

{\bf Light and strange baryons}: The left of Fig. \ref{lcbaryons} is a summary plot of lattice results 
for the octet and the decuplet light baryon masses at physical pion mass. The results from ``BMW $N_f=2+1$'' 
\cite{Durr:2008zz} and ``ETMC $N_f=2+1+1$'' \cite{Alexandrou:2014sha} are chiral and continuum extrapolated, 
whereas the results from ``ETMC $N_f=2$'' \cite{Alexandrou:2017xwd}, ``PACS-CS $N_f=2+1$'' \cite{Aoki:2008sm}
and ``QCDSF-UKQCD $N_f=2+1$'' \cite{Bietenholz:2011qq} are at physical pion mass, but with no continuum 
extrapolations. Note that the details of the methodology considerably differ between different lattice 
calculations: {\it e.g.} the lattice ensembles being used, the fermion and the gauge field actions, the 
degree of control over the lattice systematics ({\it e.g.} arising from chiral and continuum limits), etc. 
The success of lattice investigations is reflected in the mutual agreement between their results and 
their agreement with experiments. One of the most interesting investigations of recent times is the 
precise estimation of the energy splittings in $N$, $\Sigma$, $\Xi$, $D$ and $\Xi_{cc}$ isospin multiplets 
from lattice QCD and QED computations with $N_f=1+1+1+1$ fermions by BMW collaboration \cite{Borsanyi:2014jba}. 
This was discussed in a plenary talk at Lattice 2014 \cite{Portelli:2015wna}.

Several other exploratory lattice calculations have also been performed to estimate the light as well as strange 
baryons ({\it e.g.} Refs. \cite{Bulava:2010yg,Edwards:2011jj,Edwards:2012fx,Engel:2013ig}). Note that even in the 
absence of chiral and continuum extrapolations, the results from these calculations make precise predictions 
for quantum numbers of the ground state light and strange baryon masses, {\it e.g.} for the recently discovered 
$\Omega^{\ast -}$ baryon by the Belle Collaboration \cite{Yelton:2018mag}.

{\bf Singly charmed baryons}: The right of Fig. \ref{lcbaryons} presents a summary of recent lattice results for 
the masses of singly charmed baryons. The experimental masses are shown as gray horizontal lines. The results 
from ``ILGTI `13-`18'' \cite{Basak:2012py,Basak:2013oya,Basak:2014kma,Mathur:2018rwu}, ``Briceno {\it et.al.} `12''
\cite{Briceno:2012wt} and ``Brown {\it et. al.} `14'' \cite{Brown:2014ena} are based on mixed action calculations 
and are chiral and continuum extrapolated to the physical limits. ``PACS-CS `13'' \cite{Namekawa:2013vu} and 
``ETMC `17'' \cite{Alexandrou:2017xwd} refer to the results calculated at physical pion mass, whereas ``RQCD `15''
\cite{Bali:2015lka} refers to chiral extrapolated results, all at a single lattice spacing. ``TWQCD `17'' 
\cite{Chen:2017kxr} and ``D\"urr {\it et. al.} `12'' \cite{Durr:2012dw} are exploratory investigations on single 
lattice QCD ensembles at heavier than physical pion masses. ``HSC `15'' refers to the results from an exploratory 
study of excited charm baryon spectrum on an anisotropic ensemble with $m_{\pi}=391$ MeV \cite{Padmanath:2015bra}. 
It is very evident from the figure that there is a good overall agreement between all the lattice estimates and also 
with the respective experimental masses. Note that lattice results for heavy quarks are expected to be severely 
affected by discretization effects. Hadrons with a larger number of valence heavy quarks are expected to be affected 
with larger discretization effects. Some of these calculations utilize novel techniques like Fermilab approach to 
tune the heavy quark masses, and mass differences and dimensionless mass ratios to perform extrapolations. These 
procedures are expected to remove the leading discretization effects and to provide good control in continuum 
extrapolations. In this sense, the agreement between different lattice estimates and with experiment implies smaller
discretization effects in comparison with the statistical uncertainties for the lattice fermion actions used for 
charm quarks at the lattice spacings utilized. This also gives confidence in making robust and reliable predictions 
for the doubly and the triply heavy baryons. 

\begin{figure}
\includegraphics[height=4.5cm,width=7.2cm]{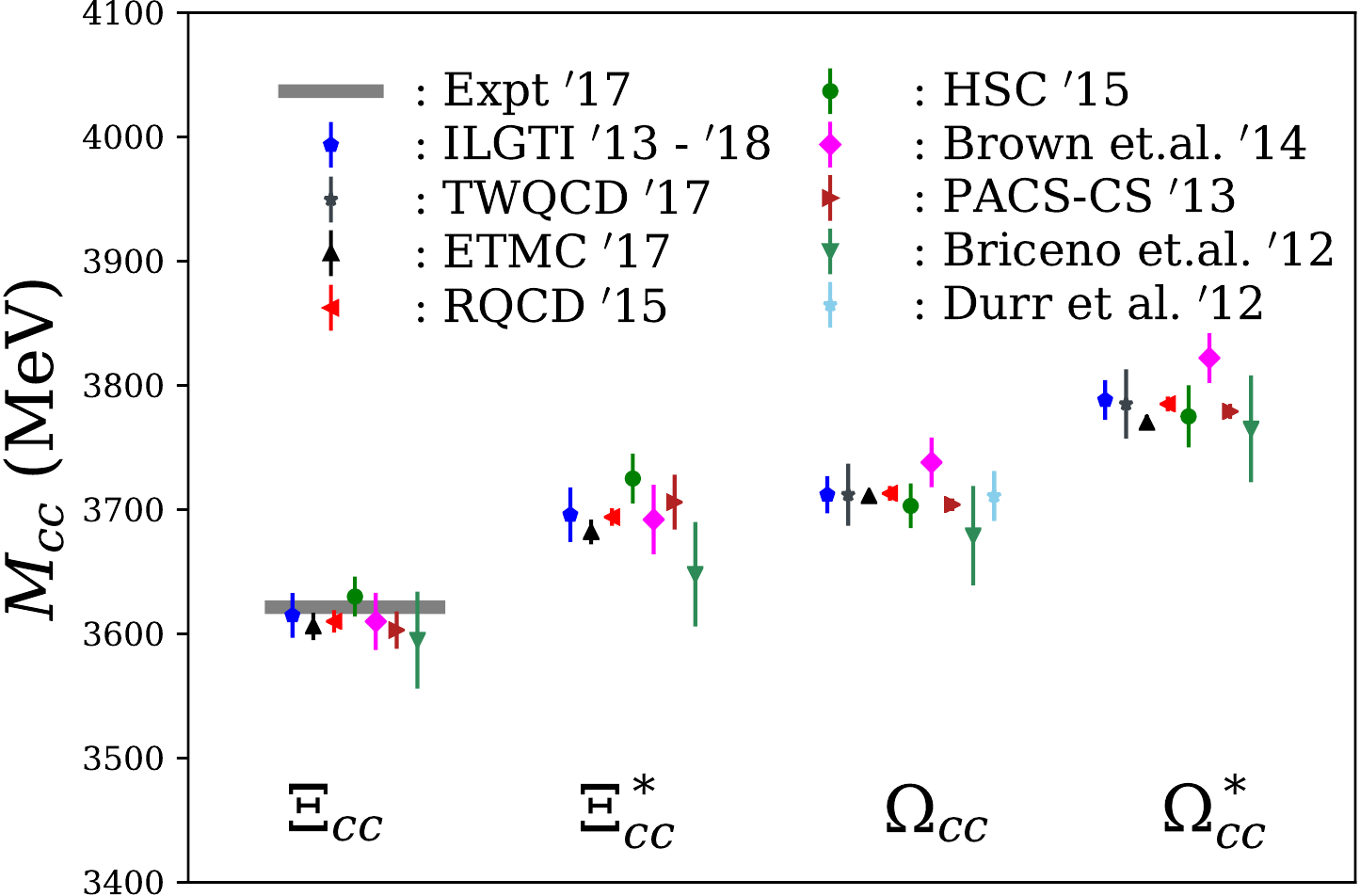}
\includegraphics[height=4.5cm,width=7.2cm]{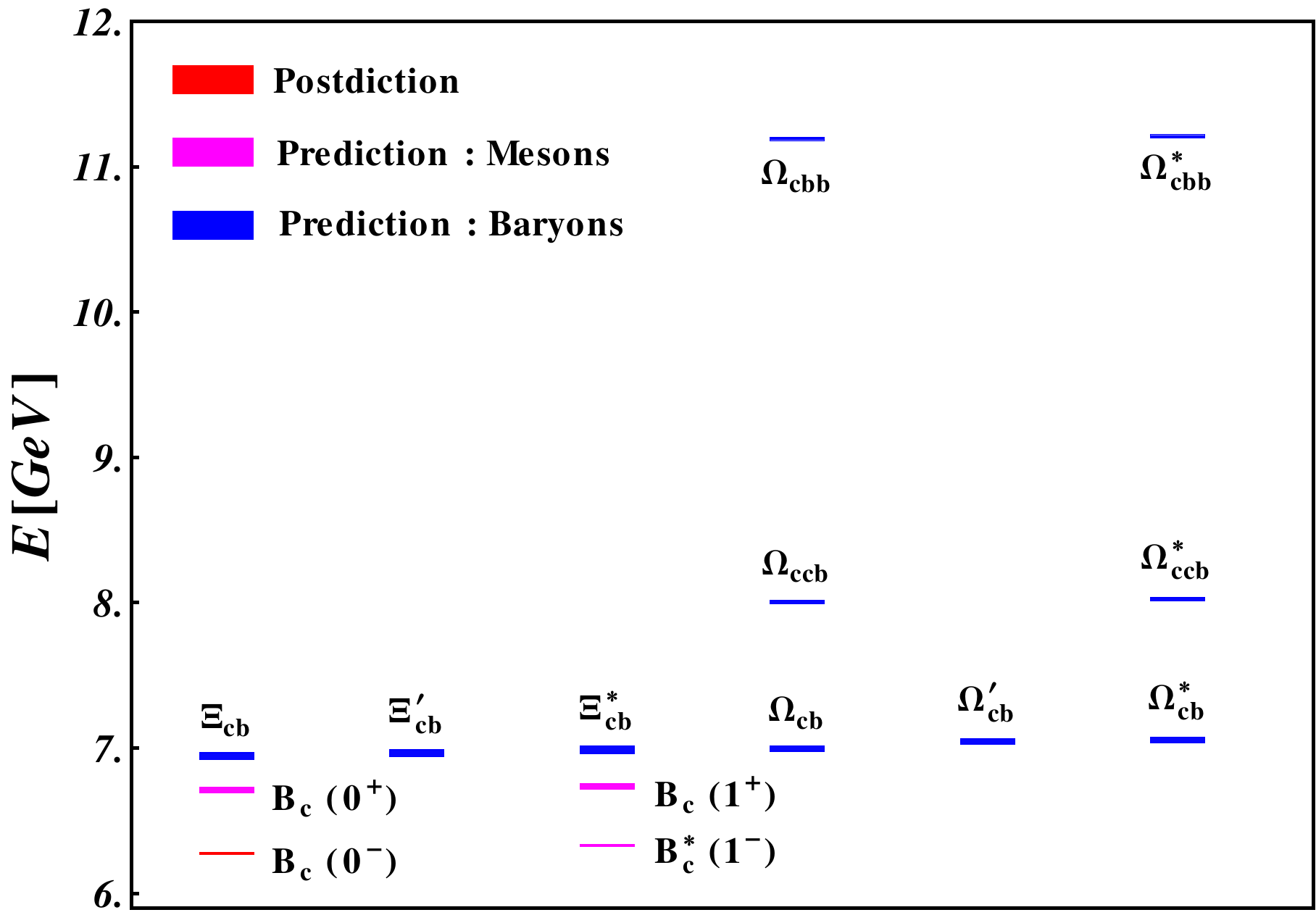}
\caption{Left: Lattice predictions for the masses of the doubly charm baryons. The experimental mass for $\Xi_{cc}(1/2^+)$
as determined by LHCb \cite{Aaij:2018gfl} is shown as a horizontal band. Right: Lattice estimates for charmed-bottom hadron
masses as determined in Ref. \cite{Mathur:2018epb}.}\label{hbaryons}
\end{figure}

{\bf Doubly charmed baryons}: In the left of Fig. \ref{hbaryons}, lattice predictions for the ground state doubly 
charmed baryon masses are summarized. It is evident from the figure that these estimates for the only known doubly heavy 
system $\Xi_{cc}(1/2^+)$ show good agreement with each other. They are all consistent with the experimental mass as 
determined by the LHCb Collaboration \cite{Aaij:2018gfl}. Note that all these lattice estimates predate the LHCb 
discovery and hence were predictions for this state. This clearly demonstrates the potential of lattice QCD techniques 
to make reliable predictions in the heavy baryon sector\footnote{Several successful lattice predictions exist also in the 
heavy meson sector ({\it c.f.} Ref. \cite{Bazavov:2009bb,Dowdall:2012ab}).}. Not all the results have been estimated 
after a continuum extrapolation. This indicates that the cut-off uncertainties, that are expected to be severe in 
the heavy hadron observables, are small. In this figure, the results from a few early quenched lattice calculations 
\cite{Lewis:2001iz,Mathur:2002ce,Flynn:2003vz} and a dynamical calculation \cite{Liu:2009jc} on the ground state 
heavy baryon masses, which are also in good agreement with results presented in Fig. \ref{hbaryons}, are omitted. 

All the lattice predictions for the mass of $\Xi_{cc}(1/2^+)$, shown in Fig. \ref{hbaryons}(left), lie $\sim 100$ MeV above 
the SELEX measurement for the mass of a doubly charmed baryon (3519(1) MeV) \cite{Mattson:2002vu}. As pointed out earlier, 
a precision determination of the energy splittings in $N$ and $\Xi_{cc}$ isospin multiplets from lattice QCD and QED 
computations with $N_f=1+1+1+1$ fermions was performed by BMW collaboration \cite{Borsanyi:2014jba}. In this calculation,
they postdict the neutron-proton mass splittings with an accuracy of 0.3 MeV. The lattice prediction for the energy 
splitting between the isospin partners of $\Xi_{cc}(1/2^+)$ from this calculation is $2.16(11)(17)$ MeV. This excludes 
the possibility that the SELEX measured doubly charmed baryon candidate to be the isospin partner of $\Xi_{cc}^+(3621.4\rm{~MeV})$. 

{\bf Charmed bottom baryons}: In the right of Fig. \ref{hbaryons}, lattice estimates for the masses of hadrons with at 
least one charm and one bottom quark are presented. These results are from mixed action calculations using overlap 
fermions for quark masses up to charm and a non-relativistic QCD formulation for bottom studied on $N_f=2+1+1$ HISQ 
fermion MILC ensembles \cite{Mathur:2018epb}. This investigation was carried out on three ensembles with different 
lattice spacings to achieve good control over the discretization effects. Utilizing the energy splittings and 
dimensionless mass ratios, the authors perform controlled chiral and continuum extrapolations to obtain reliable 
predictions for many yet to be discovered charmed-bottom hadrons. The lattice postdiction for the only discovered 
charmed-bottom hadron $B_c$ meson is found to be in good agreement with the experimental mass. Note that the 
experimental mass of this $B_c$ meson was originally found to be in agreement with the lattice prediction 
\cite{Dowdall:2012ab}. The predictions in Fig. \ref{hbaryons}(right) for ground state $B_c$ meson masses are also 
in agreement with the predictions in Ref. \cite{Dowdall:2012ab}. The mass estimates for charmed-bottom baryons are 
also found to be in good agreement with the only existing previous dynamical calculation \cite{Brown:2014ena} 
performed on RBC-UKQCD ensembles with two different lattice spacings. 

All the results presented above are estimated within the single hadron approach, where only three quark interpolators are 
considered in the analysis and the effects of any nearby strong decay thresholds are neglected. This is justified for most 
baryons discussed above, considering the fact that they are deeply below the respective lowest strong decay thresholds. 
However, this approach is questionable for baryons like $\Delta$, $\Sigma^*$, $\Xi^*$, etc., which are resonances and can 
decay into one or more strong decay modes. Hence attributing bare lattice energy levels to the resonance energies is not 
appropriate and they require a rigorous finite volume analysis as mentioned in Section \ref{methodology}. Recent lattice 
calculations in this regard will be part of the discussion in Sections \ref{mesres} and \ref{barres}. 

\section{Excited hadron spectroscopy}\label{excited}

The first step in performing a rigorous finite volume treatment on the lattice is to reliably extract the discrete 
energy spectrum. As mentioned previously, a standard practice these days is to evaluate correlation matrices (Eq. 
\ref{eq:2-1}) for a basis of interpolating operators and solve the GEVP (Eq. \ref{eq:2-2}) to extract the excited 
state information. A procedure that is developed and followed by HSC to build meson interpolator basis on the lattice 
has been quite successful in extracting multiple excited states and reliably identify their quantum numbers 
\cite{Liao:2002rj,Dudek:2007wv,Dudek:2010wm,Thomas:2011rh}. This procedure has been utilized extensively by HSC in 
their lattice investigations for light mesons, charmed mesons and charmonia \cite{Dudek:2009qf,Thomas:2011rh,Liu:2012ze,
Moir:2013ub,Cheung:2016bym}. More recently RQCD collaboration has also started practicing this formulation to study 
the charmonium spectrum \cite{Padmanath:2018tuc}. Lattice calculations utilizing other interpolator bases in extracting 
excited state spectrum for heavy quarkonium as well as heavy-light mesons also exist in the literature \cite{Bali:2011rd,
Mohler:2012na,Wurtz:2015mqa}.

\begin{figure}
\includegraphics[height=4.8cm,width=7.2cm]{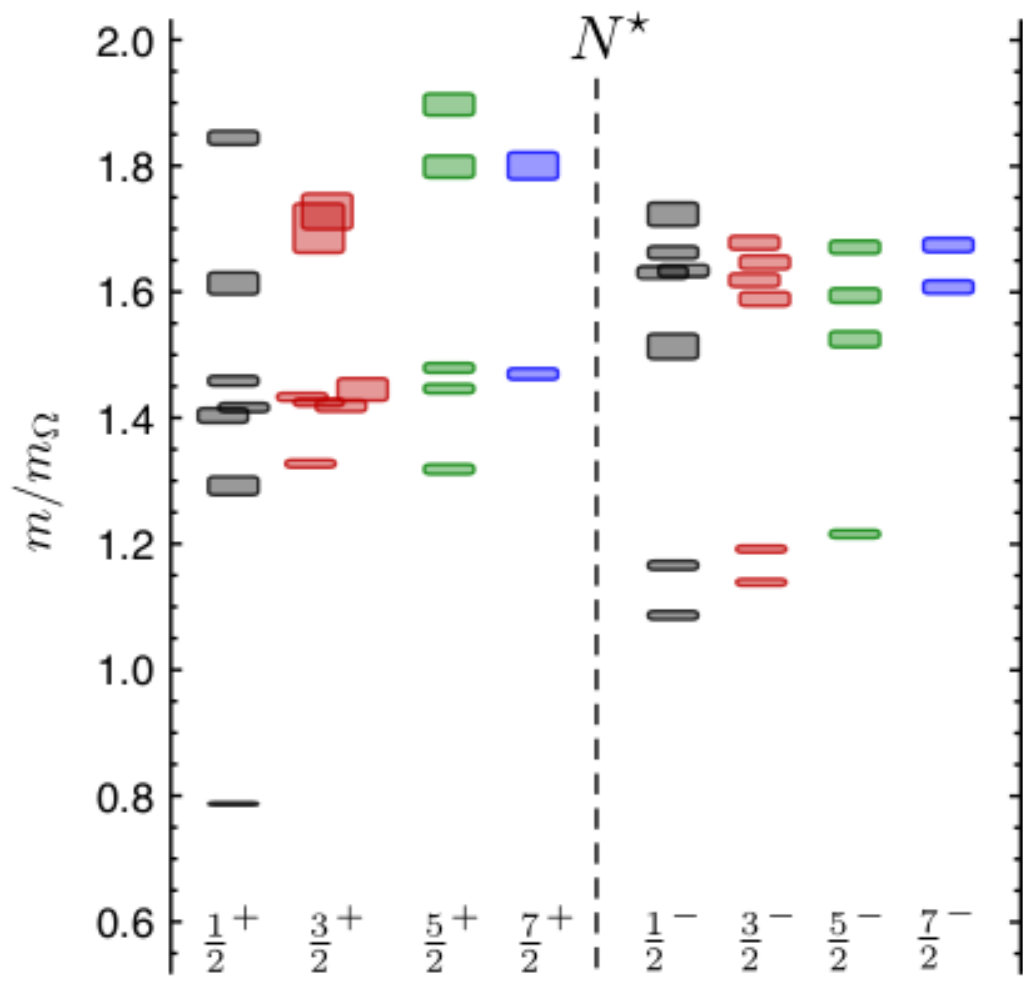}
\includegraphics[height=4.8cm,width=7.6cm]{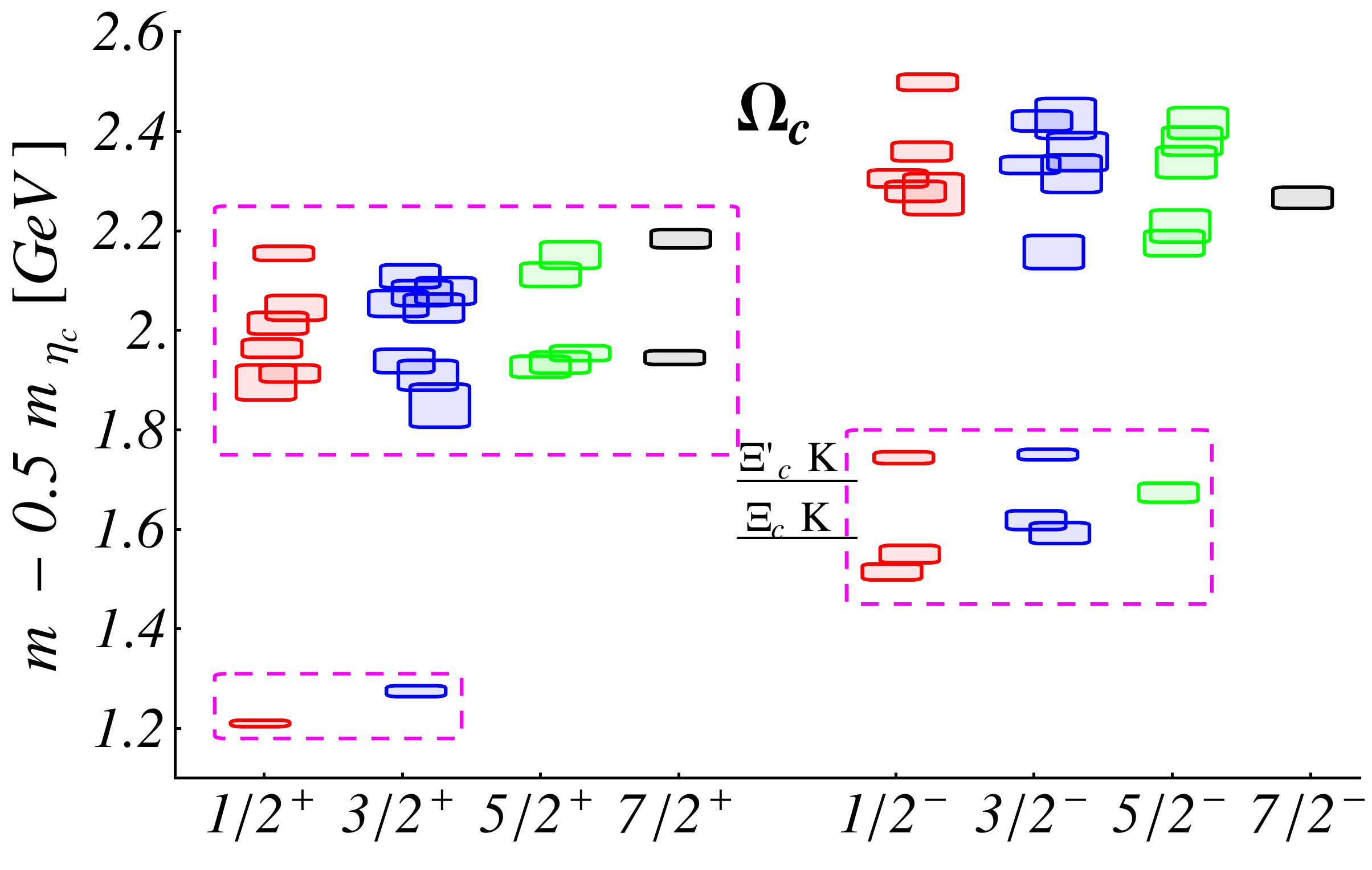}
\caption{Spin identified excited state spectra of nucleon (left) and $\Omega_c$ baryon (right) as determined in Refs. \cite{Edwards:2011jj} and \cite{Padmanath:2017lng}, respectively.}\label{excitedbaryons}
\end{figure}

{\bf Excited baryons}: An equivalent procedure to systematically build baryon interpolators has been developed by 
the LHP Collaboration more than a decade ago \cite{Basak:2005ir,Basak:2005aq}. Early calculations following these 
interpolators have also been reported in Ref. \cite{Bulava:2009jb}. Over the past years, HSC has realized these 
interpolators and studied light and strange baryons \cite{Bulava:2010yg,Edwards:2011jj,Edwards:2012fx} as well as 
charm baryons with one, two and three valence charm quarks \cite{Padmanath:2013zfa,Padmanath:2015jea,Padmanath:2015bra,
Padmanath:2017lng}. Fig. \ref{excitedbaryons} shows the spin identified excited spectra of nucleon (left) and 
$\Omega_c$ baryon (right) as determined in Refs. \cite{Edwards:2011jj} and \cite{Padmanath:2017lng}, respectively. 
Similar lattice investigation to extract $\Omega_{bbb}$ baryons was reported in Ref. \cite{Meinel:2012qz}.  

All lattice calculations discussed in this section follow the single hadron approach and assume that the lattice 
estimates for the mass of resonances are correct up to the respective decay width. This approach is justified 
in determining the energy of excitations that are well below the lowest allowed strong decay threshold. Narrow 
elastic resonances can also be approximately studied within this approach. Most light hadron resonances have 
decay widths of the order of 100 MeV and hence are not appropriate to be studied in this way. However narrow 
resonances, such as the recently discovered excited $\Omega_c$ baryons by the LHCb Collaboration \cite{Aaij:2017nav},
can be studied using this approach. In what follows, we discuss such a lattice calculation that made precise 
predictions for the masses and quantum numbers of these excited $\Omega_c$ baryons.

\begin{figure}
\includegraphics[height=5.0cm,width=7.2cm]{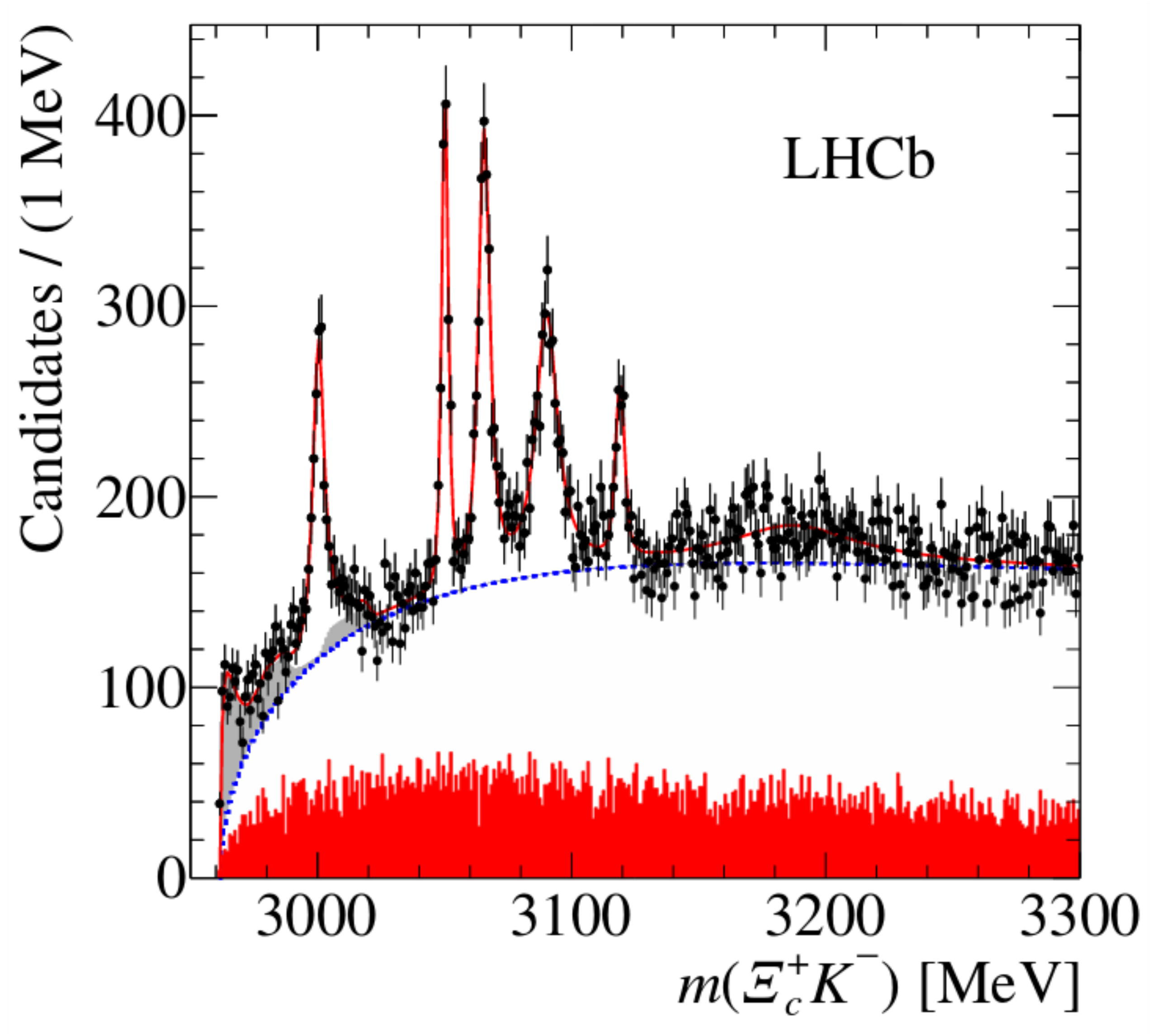}
\includegraphics[height=5.0cm,width=7.2cm]{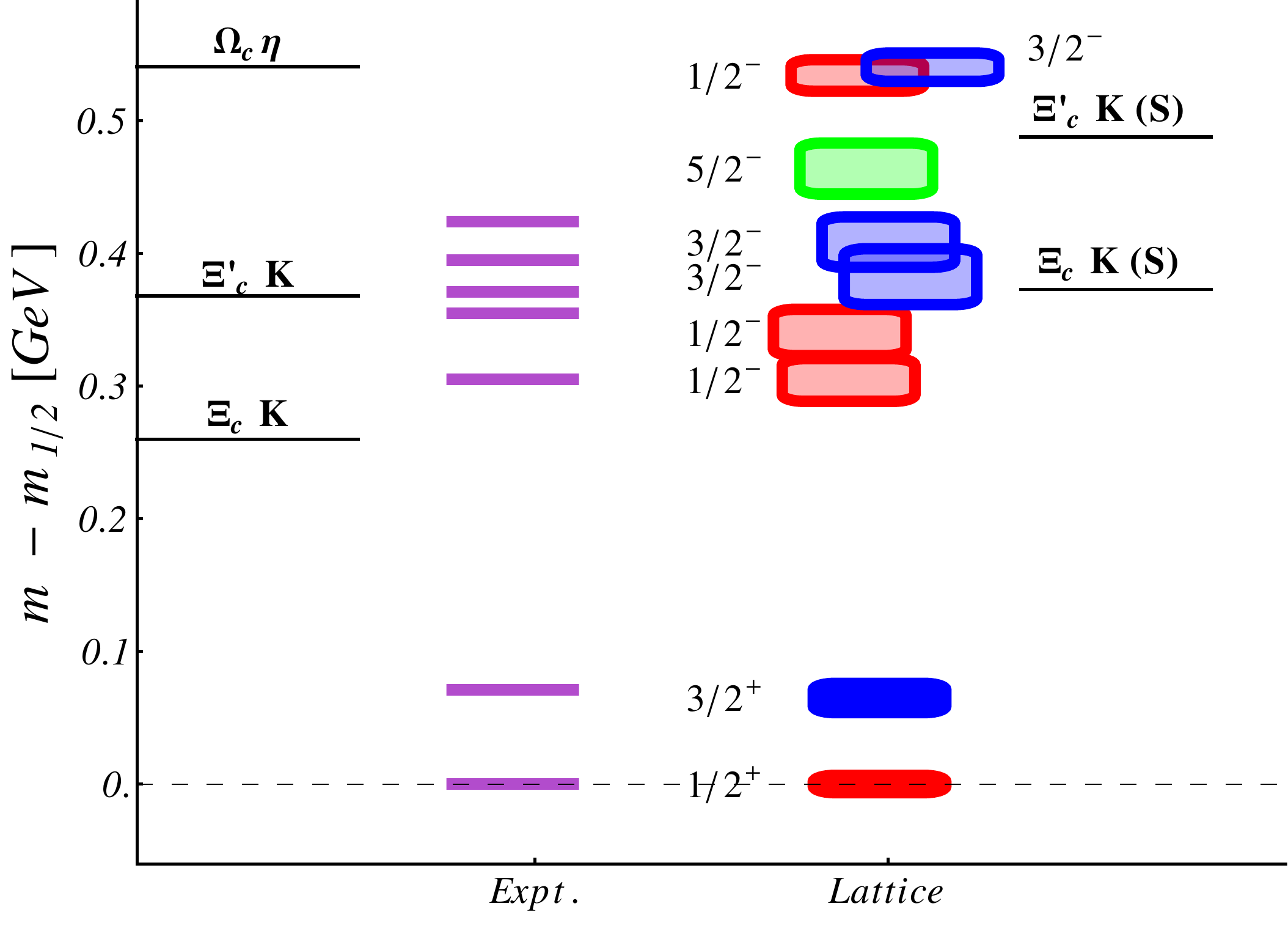}
\caption{Left: (Figure adapted from Ref. \cite{Aaij:2017nav}) The resonance structures observed in the $\Xi_c^+K^-$ 
decay mode interpreted as excited $\Omega_c$ baryons by the LHCb Collaboration. Right: Comparison of experimental 
masses of $\Omega_c$ baryons with the excited energy spectra on the lattice and their quantum number assignments. 
Nearby scattering thresholds are shown on the left and the relevant non-interacting level positions on the lattice 
are shown on the right as black horizontal lines. }\label{Ocbaryons}
\end{figure}

{\bf Excited $\Omega_c$ baryons}: The left of Fig. \ref{Ocbaryons} shows the event distribution in the $\Xi_c^+K^-$ 
decay channel displaying the five narrow peaks discovered and interpreted as excited $\Omega_c$ baryons by the LHCb 
Collaboration \cite{Aaij:2017nav}. The resonance structures can be seen to be quite narrow ($\lesssim 10$ 
MeV). Four out of these five resonances have been later confirmed using $e^+e^-$ collision data from Belle 
\cite{Yelton:2017qxg}. In the right of Fig. \ref{Ocbaryons}, a comparison of the masses of the seven experimentally 
known $\Omega_c$ baryons (indicated by the horizontal magenta lines) with the lowest nine $\Omega_c$ baryons as 
extracted in Ref. \cite{Padmanath:2017lng} is made. This calculation is performed on an $N_f=2+1$ anisotropic clover 
ensemble with a pion mass of 391 MeV and a physical spatial volume $(1.9\rm{~fm})^3$ generated by HSC, and ignores 
any strong decay modes. The results from this calculation correctly postdict the mass of $\Omega_c(1/2^+)$ baryon and the 
hyperfine splitting in the ground state $\Omega_c$ baryons ($E_{3/2^+}-E_{1/2^+}$). Note that the hyperfine splittings 
are generally observed to be quite sensitive to discretization uncertainties. The agreement between the lattice 
and experiment indicates such uncertainties on $\Omega_c$ baryons in this lattice setup are small. Lattice also 
predicts five states as shown in the figure, with quantum numbers ``$1/2^-,1/2^-,3/2^-,3/2^-\rm{~and~}5/2^-$'' in 
the region of experimental discovery. Note that these results were reported in Lattice 2014 \cite{Padmanath:2014bxa} 
as well as in Charm 2013, 2015 \cite{Padmanath:2013bla,Padmanath:2015bra} and hence predate the LHCb discovery. An 
immediate extension to this calculation would be to include interpolators that are related to the nearby non-interacting 
baryon-meson levels and to perform a finite volume analysis. Such calculations including baryon-meson interpolators 
in the analysis are in their early stage of development and will be discussed in Section \ref{barres}. More lattice 
calculations of the excited baryon spectrum will be highly appreciated by the scientific community, anticipating 
the discovery of many more baryons in experiments like LHCb and Belle. 

\section{Meson resonances on the lattice}\label{mesres}

The majority of hadrons are resonances and can decay via strong interactions.  Most resonances have large decay widths 
$\mathcal{O}(100 MeV)$ and can decay into different sets of hadronic final states. The studies of hadronic resonances 
on the lattice demand a finite volume treatment. A good way to start such calculations is to explore the 
easiest cases of elastic resonances and shallow bound states. Gradually relaxing various simplifying approximations, 
one may investigate more complicated scenarios like the effects of inelastic thresholds, coupled channel scattering 
and so on. 

{\bf Elastic scattering in light mesons}: The pseudoscalar-pseudoscalar elastic scattering in the light and strange 
meson sector are two widely performed benchmark calculations among different lattice groups. There have been many lattice 
investigations of the $\rho$ (vector) meson in $p$-wave $\pi\pi$ scattering \cite{Aoki:2007rd,Jansen:2009hr,Aoki:2011yj,
Lang:2011mn,Pelissier:2012pi,Dudek:2012xn,Wilson:2015dqa,Bali:2015gji,Guo:2016zos,Fu:2016itp,Alexandrou:2017mpi}. Some 
recent calculations in this channel have also been reported during this meeting. The coupling $g_{\rho.\pi\pi}$ has been 
generally observed to be independent of the pion mass in the elastic regime. A summary of different lattice results is 
made in Fig. 12 of Ref. \cite{Alexandrou:2017mpi}. It is seen that a dimensionless ratio $m_{\rho}/m_N$ is found to 
roughly indicate linear dependence with the pion mass squared leading to the experimental value in the chiral limit. 
The results from $N_f=2$ calculations are found to be scattered around the results from $N_f=2+1$ studies. A discussion 
on the pion mass dependence and the quenching effects of the strange sea on the mass of the $\rho$ meson can be found 
in Ref. \cite{Alexandrou:2017mpi}. The $K^*$ meson is also studied in many lattice calculations through $p$-wave $K\pi$ 
scattering \cite{Prelovsek:2013ela,Wilson:2014cna,Dudek:2014qha,Bali:2015gji,Brett:2018jqw}. Among the calculations cited 
above, two unique calculations are those performed by HSC a few years back \cite{Wilson:2015dqa,Wilson:2014cna,
Dudek:2014qha}. In these articles, the authors have studied the effects of the inelastic threshold by performing a coupled 
channel finite volume analysis.

\begin{figure}
\centering
\includegraphics[height=5.8cm,width=15cm]{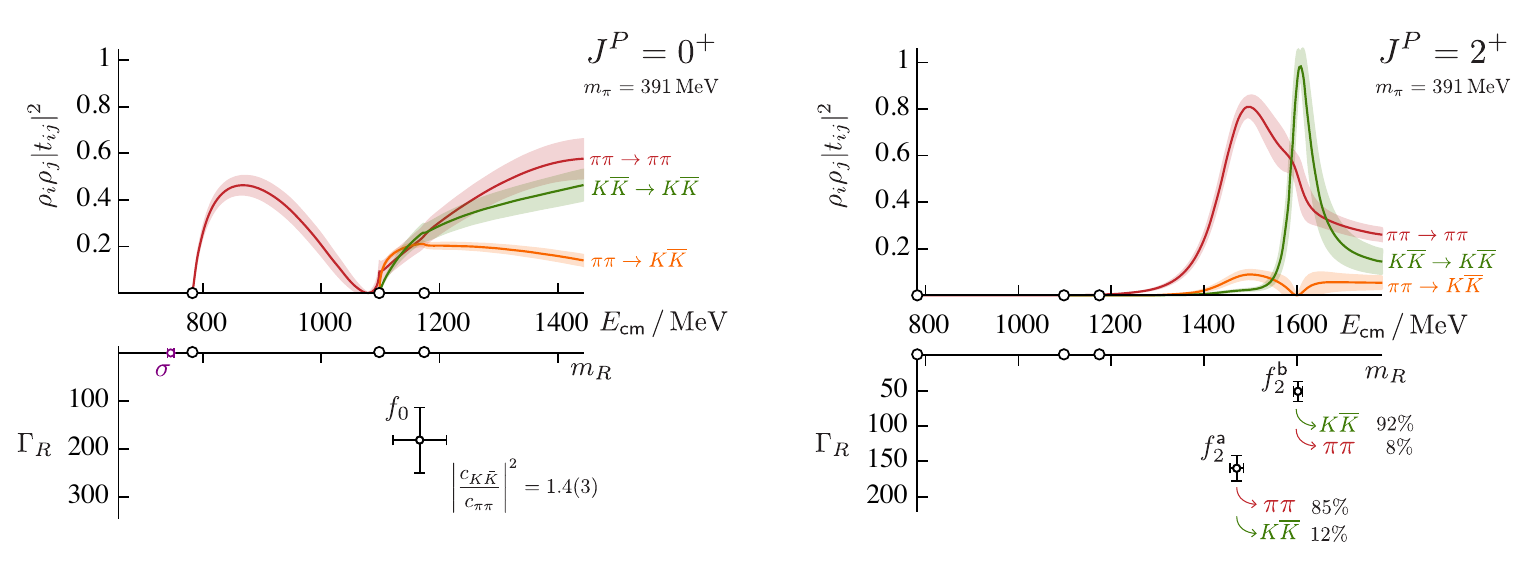}
\caption{ (Figure adapted from Ref. \cite{Briceno:2017qmb}) Coupled $\pi\pi$-$K\bar K$ amplitudes for isoscalar-scalar (left) and 
isoscalar-tensor (right) mesons. The three black circles in the real axis are the scattering thresholds corresponding to $\pi\pi$, 
$K\bar K$ and $\eta\eta$. In both the channels, $\eta\eta$ is found to be approximately decoupled. Pole singularities determining 
the features of the coupled channel scattering amplitudes are shown along with their uncertainties that include the variation 
in amplitude parameterizations.}
\label{hscscalar}
\end{figure}

{\bf Coupled channel scattering in light mesons}: Recently HSC has performed a coupled channel ($\pi\pi$, $K\bar K$, 
$\eta\eta$) investigation in the isoscalar-scalar channel, where the $\sigma$ meson and the $f_0(980)$ appear 
\cite{Briceno:2017qmb}. This calculation is an extension of their earlier investigation with $s$-wave $\pi\pi$ elastic 
scattering \cite{Dudek:2013yja} and also includes an investigation of $d$-wave couple channel scattering. The summary 
of their results is as shown in Fig. \ref{hscscalar}. Owing to the heavy pion mass of 391 MeV, the $\sigma$ meson is 
a stable bound state in their setup. As in experiment, the $f_0(980)$ features as a dip in the $\pi\pi$ cross-section 
close to the $K\bar K$ threshold. Two resonance peaks observed in the $d$-wave scattering amplitudes are argued to be
related to $f_2(1270)$ and $f_2'(1525)$, with the lighter peak decaying predominantly to $\pi\pi$ and the heavier peak 
to $K\bar K$. There also exist lattice efforts employing lighter pion masses to investigate this channel from HSC 
\cite{Briceno:2016mjc} and others \cite{Guo:2018zss}. In another recent article, HSC has demonstrated the application of 
their techniques and the analysis of dynamically coupled partial waves in isospin-2 $\rho\pi$ scattering \cite{Woss:2018irj}.  

\begin{figure}
\includegraphics[height=4.5cm,width=7.5cm]{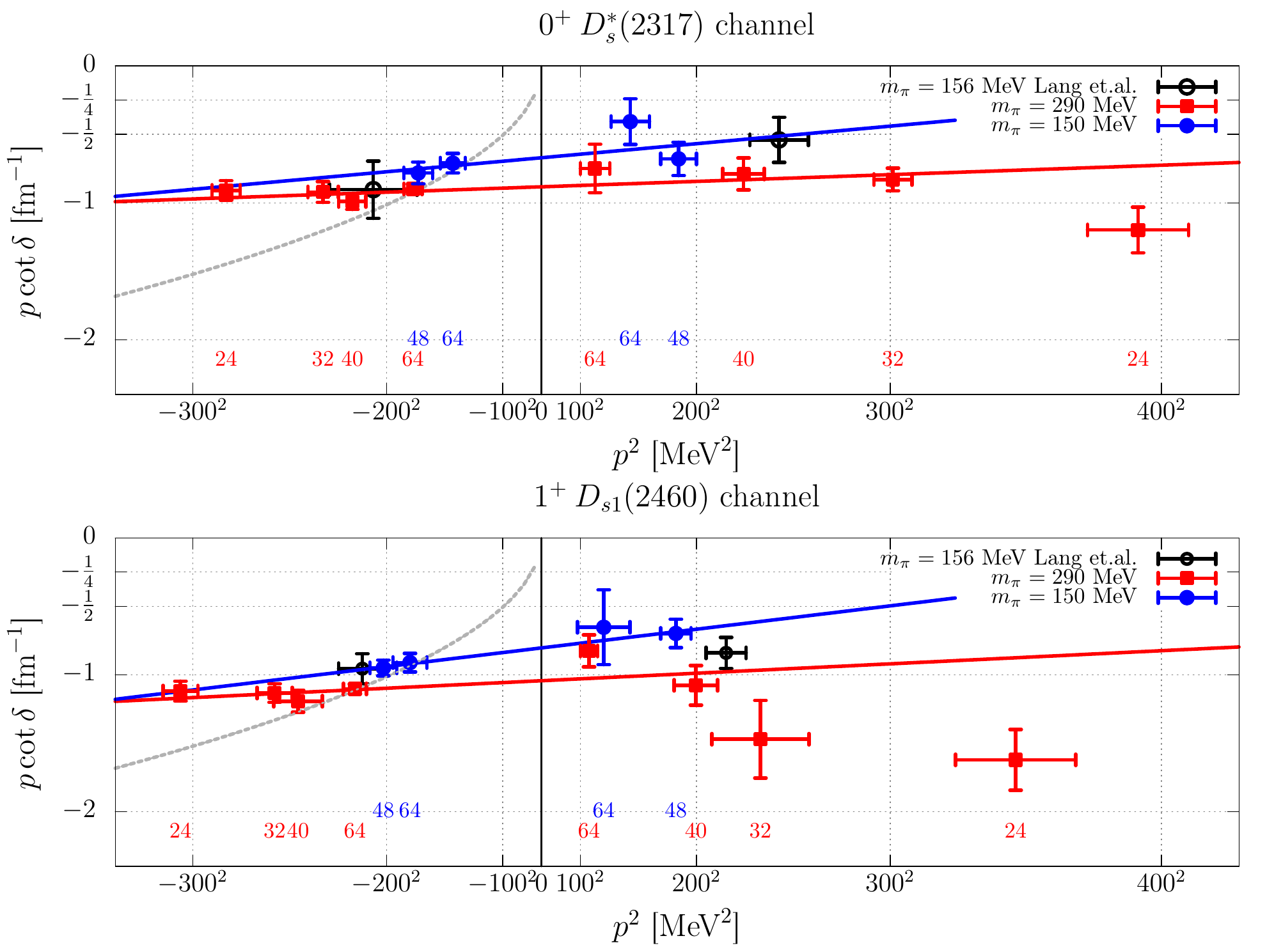}
\includegraphics[height=4.5cm,width=7.1cm]{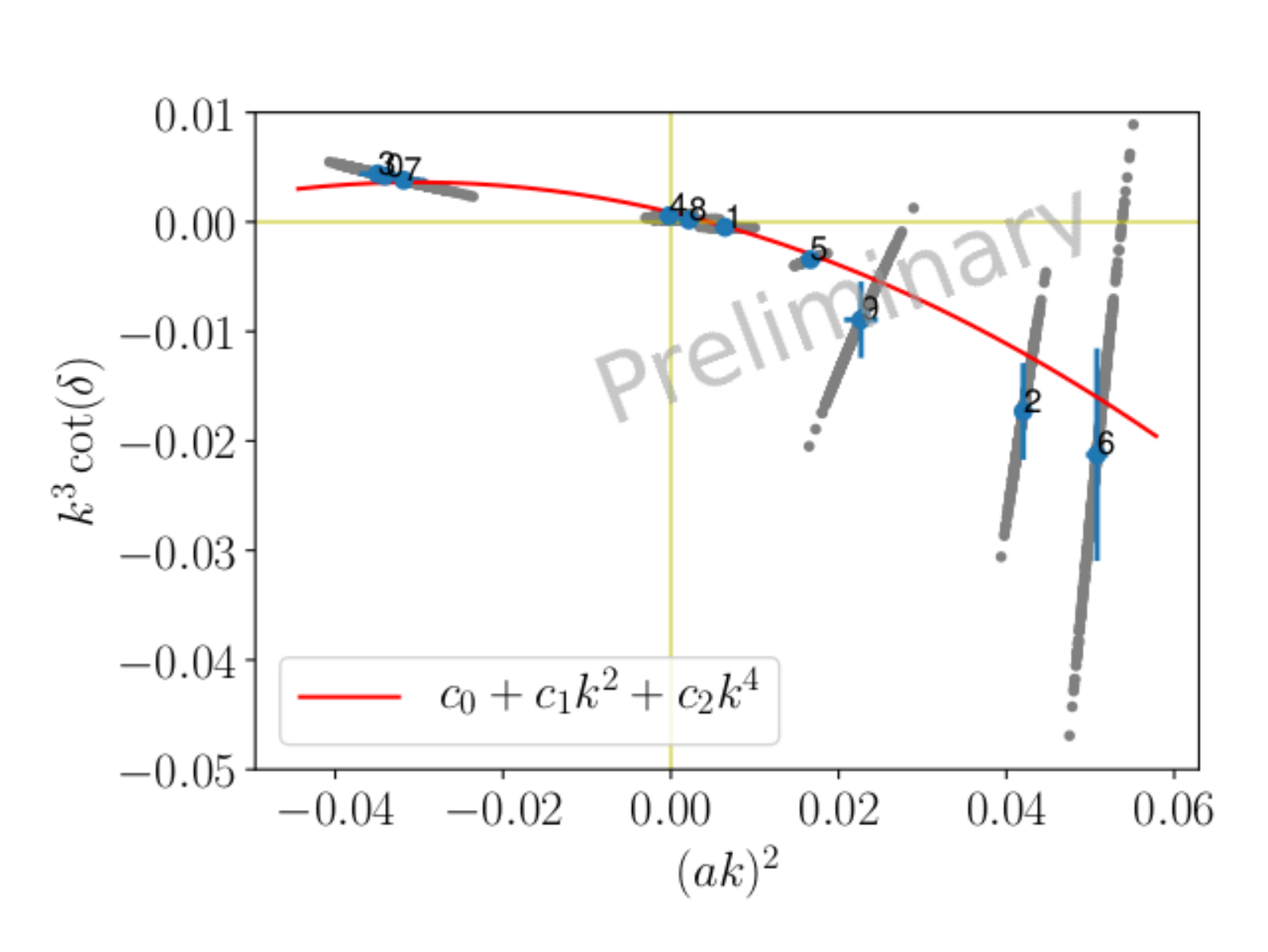}
\caption{Left: (Figure adapted from Ref. \cite{Bali:2017pdv}) $p~cot\delta$ as a function of $p^2$ for the scalar ($0^+$, top) 
and vector ($1^+$, bottom) heavy-strange mesons. The red and blue lines are linear fits to the data determining the bound state 
position from the intersection with the bound state constraint curve ($ip=-\sqrt{-p^2}$) shown in gray dashed curves. Right: 
$k^3cot\delta$ as a function of $k^2$ for vector ($1^-$) charmonium mesons from a calculation by the RQCD Collaboration. The 
red curve indicates a fit to the phase shift observed with a quartic fit form to describe a bound state and a resonance.
}\label{Dsbarcc}
\end{figure}

{\bf Charmed mesons}: In contrast to phenomenological expectations, the scalar $0^+$ and axial-vector $1^+$ ground states 
of heavy-strange mesons are found to be narrow and below the scattering thresholds $KD$ and $KD^*$, respectively. A first 
calculation involving elastic scattering was performed in Ref. \cite{Mohler:2013rwa} on a PACS-CS ensemble with physical 
spatial volume $(2.9\rm{~fm})^3$ and near to physical pion mass. In a recent calculation by the RQCD Collaboration, the 
authors study these mesons utilizing six lattice QCD ensembles with $N_f=2$ non-perturbatively $\mathcal{O}(a)$ improved 
Wilson sea quarks at $a=0.07$ fm, covering several spatial volumes with $L$ as large as $4.5$ fm and two different pion 
masses (290 MeV and 150 MeV). Performing a phase shift analysis and employing the effective range approximation, they 
determine the bound state masses and the coupling with the respective thresholds. In the left of Fig. \ref{Dsbarcc}, linear 
fits to the phase shift data in the scalar as well as the axial-vector channel to describe the $D_{s0}^*(2317)$ and the 
$D_{s1}(2460)$ are shown. They also extract lattice levels in the axial-vector channel related to the $D_{s1}(2536)$ 
resonance, which are resolved sufficiently well using only the $\bar cs$ interpolators. The energies of these three states 
are found to be very sensitive to the pion mass (changes $\sim$30 MeV from 290 MeV to 150 MeV) in comparison with the 
changes in the energies for ground state pseudoscalar and vector $D_s$ mesons, which are 3 and 7 MeV, respectively. 
In addition to masses, they also determine the weak decay constants $f^{0^+}_V$ and $f^{1^+}_A$ of the $D_s$ mesons. 

{\bf Excited charmonium}: In the charmonium spectrum below 4 GeV, the vector channel has two bound states $J/\psi$, 
$\psi(2S)$ and a resonance $\psi(3770)$, whereas in the scalar channel there is a bound state $\chi_{c0}(1P)$ 
and a recently discovered resonance $\chi_{c0}(2P)$ \cite{Chilikin:2017evr}. The scalar channel is interesting due to 
the presence of another candidate $X(3915)$, for which the quantum numbers are not yet known, but are expected to be 
either $0^{++}$ or $2^{++}$. There has been only one previous calculation of these channels within the elastic 
scattering of $D\bar D$ performed in the rest frame \cite{Lang:2015sba}. Recent efforts by the RQCD Collaboration in 
studying the low lying resonant spectra in scalar and vector charmonia in the moving frames were reported at this 
meeting. Using two ensembles with $m_{\pi}\sim280$ MeV and $m_K\sim 467$ MeV, and with spatial extents 
$L\sim2$ fm and $L\sim2.7$ fm, they investigate scattering amplitudes in the vector and scalar charmonium channels up 
to an energy of 4 GeV. In the right of Fig. \ref{Dsbarcc}, the phase shifts as a function of the momentum squared 
for the vector charmonium channel as determined by the RQCD Collaboration are presented. 
  
\section{Baryon resonances on the lattice}\label{barres}

In contrast to the meson sector, baryon resonances have received very little attention, largely due to the computational 
challenges. Evaluation of a large number of Wick contractions possibly involving also annihilation diagrams, large 
computational and storage requirements owing to increase in the number of valence quarks and exponential degradation 
of signal-to-noise ratio demand a humongous amount of computational resources. In addition to all these, the non-zero 
spin of baryons complicates the phase shift analysis. In the physical pion mass limit, more challenges appear with 
the opening of new scattering thresholds, including those involving three or more hadron scattering. Over the past 
two decades several lattice calculations have been performed of excited light as well as strange baryons employing 
three-quark interpolators and following a single hadron approach (see review Ref. \cite{Liu:2016rwa}). Many of these 
calculations are quite remarkable considering the lattice technologies available then. New lattice technologies to 
compute all-to-all quark propagators \cite{Peardon:2009gh,Morningstar:2011ka} and various theoretical developments 
following L\"uscher's finite volume method promise precise determination of the finite volume spectrum and a reliable 
procedure to extract the resonance information therefrom. Following these procedures, there has been a calculation 
to determine the isospin-1/2 $N\pi$ scattering amplitude in the $s$-wave to describe the negative parity excitations 
$N(1535)$ and $N(1650)$ \cite{Lang:2012db}. Below, I discuss two recent calculations investigating isospin-3/2 and 
1/2 $N\pi$ scattering in $p$-wave to describe the $\Delta$ baryon and the Roper resonance ($N(1440)$), respectively.

\begin{figure}
\includegraphics[height=4.5cm,width=6.7cm]{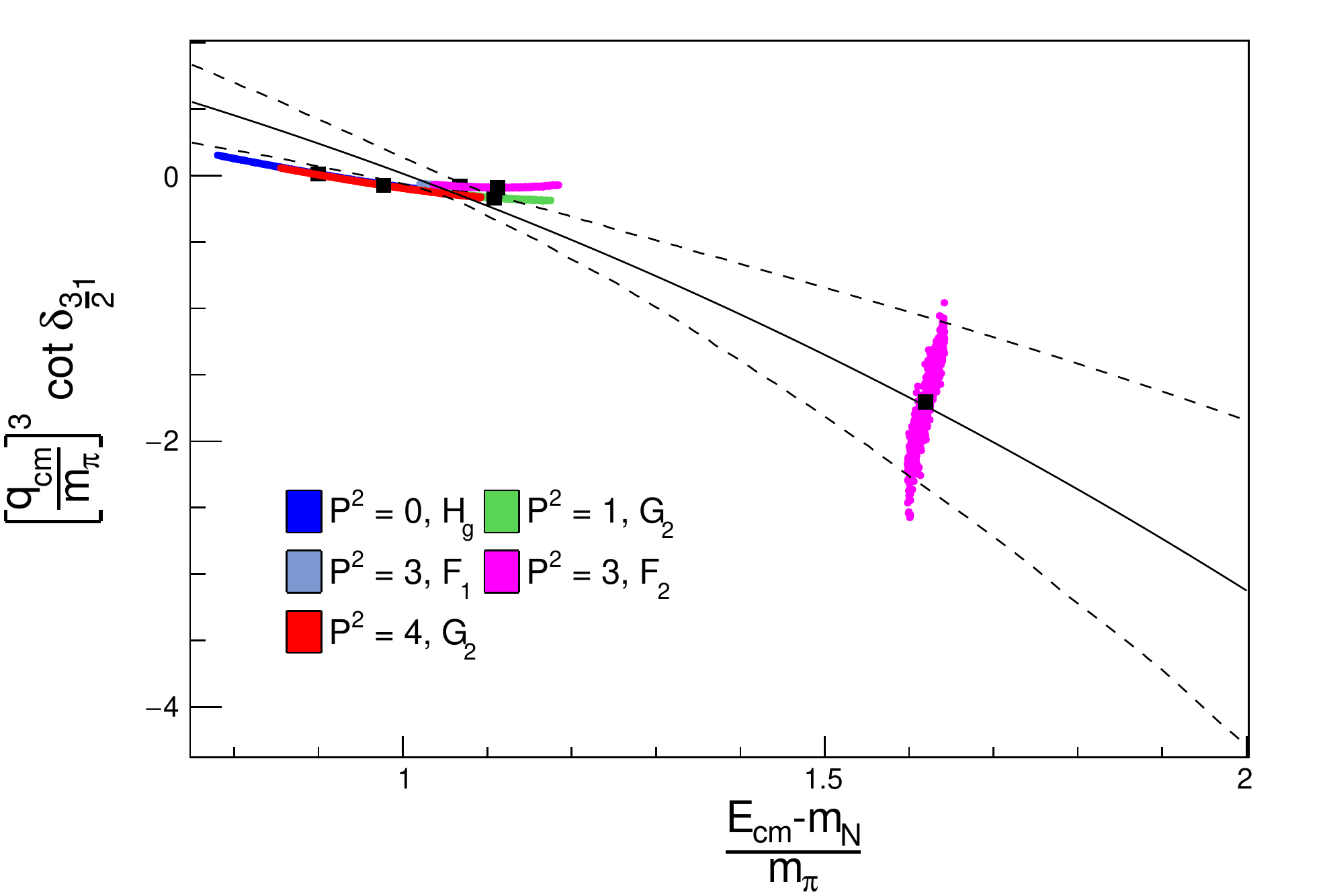}
\includegraphics[height=4.5cm,width=8.6cm]{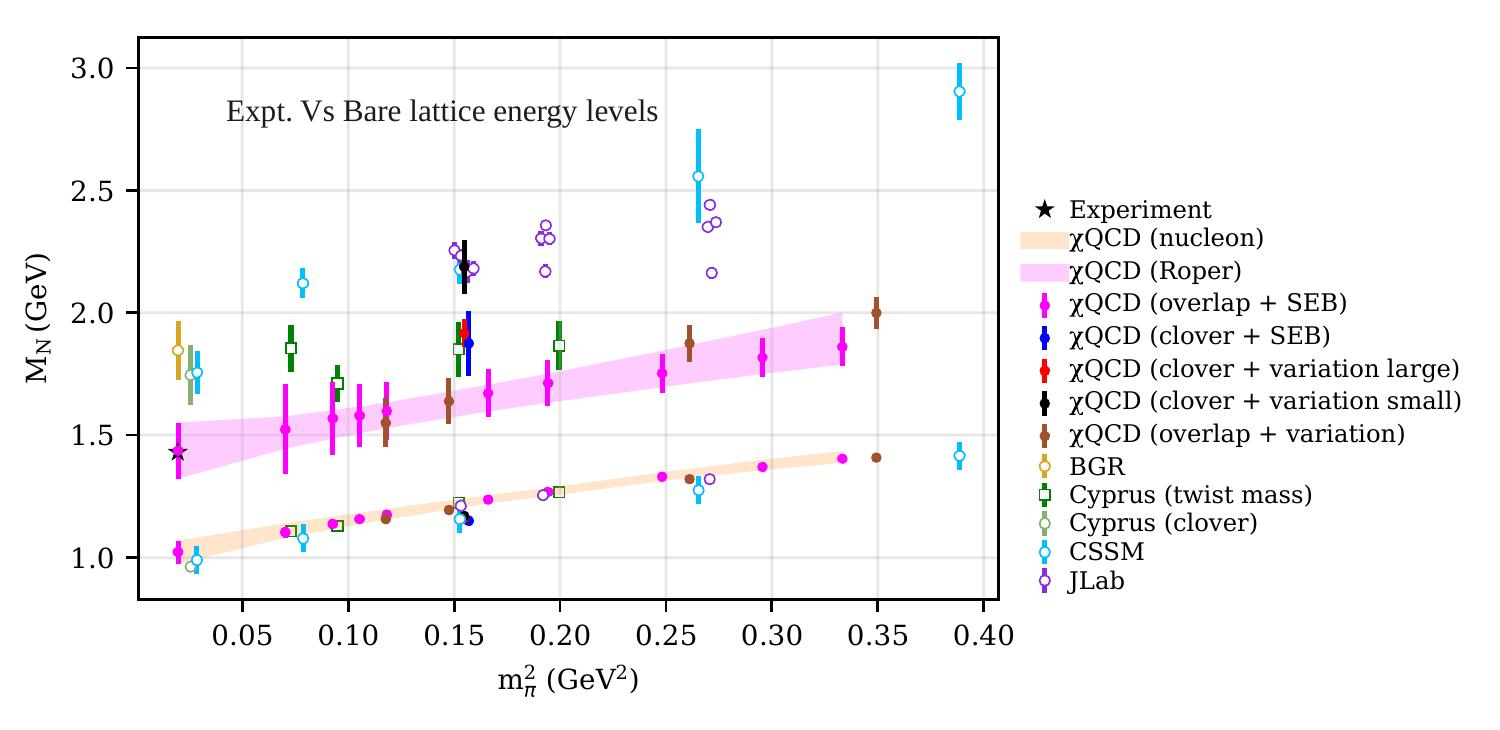}
\caption{Left: (Figure adapted from Ref. \cite{Andersen:2017una}) $\frac{q_{cm}}{m_{\pi}}^3cot\delta_{\frac32}$ as a function of 
the center of mass energy, $E_{cm}$ is shown for isospin-3/2 $N\pi$ scattering in $p$-wave. The $\Delta$ baryon 
appears in this channel. $E_{cm}$ is shown in the horizontal axis as a $(E_{cm}-m_N)/m_{\pi}$, where $m_N$ and $m_{\pi}$ are 
masses of nucleon and pion. In these units, the interval (1,2) is elastic in the $N\pi$ scattering. Right: (Figure adapted from 
Ref. \cite{Liu:2016rwa}) A summary of mass estimates for the nucleon and its first radial excitation from different lattice 
calculations. }\label{RoperDelta}
\end{figure}

{\bf $\bm{\Delta(1232)}$ baryon} is the lowest baryon resonance decaying to $N\pi$ in $p$-wave with branching fraction of 
$\sim99.4\%$ and a width $\sim 100 $ MeV. This is a good baryon candidate to be studied within an elastic approximation.  
All early calculations follow the single hadron approach. This is justified as in the ensembles with larger pion 
mass and smaller physical volumes like those used in these calculations, this baryon becomes a bound state. However, close 
to the physical pion mass and on large physical volumes determination of its resonance parameters requires a rigorous finite 
volume treatment. In Ref. \cite{Andersen:2017una}, the authors perform a finite volume analysis to extract the $\Delta(1232)$ 
baryon resonance parameters in their lattice setup. Utilizing a single lattice QCD ensemble (CLS) with $N_f=2+1$ dynamical 
flavors of Wilson clover fermions with $m_{\pi}=280$ MeV and physical lattice size $L=3.67$ fm, they observe the mass of the 
$\Delta(1232)$ baryon to be close to the $N\pi$ threshold. They find a good description of the resultant scattering amplitude 
using a Breit-Wigner shape with $m_{\Delta}=1344(20)$ MeV and $g^{BW}_{\Delta N\pi}=19.0(4.7)$. The left of Fig. 
\ref{RoperDelta} (adapted from Ref. \cite{Andersen:2017una}) shows the phase shift as a function of the center-of-mass energy. 

{\bf The Roper resonance (N(1440))} is another interesting low lying light baryon resonance. It was postulated by L. D. Roper to 
describe the $N\pi$ scattering data below 1.7 GeV \cite{Roper:1964zza}. It is the first radial excitation of the nucleon, observed 
in two hadronic final states, $N\pi$ and $N\pi\pi$ and has a total width of 350 MeV. The $N\pi\pi$ final states could be arising 
from $N\eta$, $\Delta\pi$ and $N\sigma$. This indicates the need for a coupled channel finite volume analysis, possibly including 
three hadron interpolators, to discern the Roper resonance from the lattice. In the right of Fig. \ref{RoperDelta}, the bare lattice 
energy levels ignoring possible strong decay modes for the ground and first excited state of the nucleon as a function of the 
pion mass squared (see Ref. \cite{Liu:2016rwa} for a review) are shown. It is evident that all lattice calculations yield an 
expected nucleon mass that extrapolates to the physical value in the chiral limit. However, the first excitation is consistently 
found to be close to or above 1.7 GeV even in the chiral limit, with the exception of calculations performed using chiral fermions 
\cite{Liu:2014jua,Mathur:2003zf}. Among these, only one investigation used five-quark interpolators \cite{Kiratidis:2016hda}. 
However, no levels in the energy interval (1.2,1.8) GeV are observed, including those levels related to expected scattering channels 
that are inevitable in the theory. Discussions on the disagreement between results from chiral and non-chiral actions are made in 
parallel talks during this meeting. Note that none of these calculations perform a rigorous finite volume treatment. Furthermore, 
no calculation finds any low lying levels in the energy region (1.2,1.8) GeV except for those using chiral fermions. The low lying 
first excitations in the chiral fermion calculations could be related to one (or a mixture) of the expected scattering levels 
that are inevitable in the theory.  

\begin{figure}
\centering
\includegraphics[height=4.5cm,width=6.7cm]{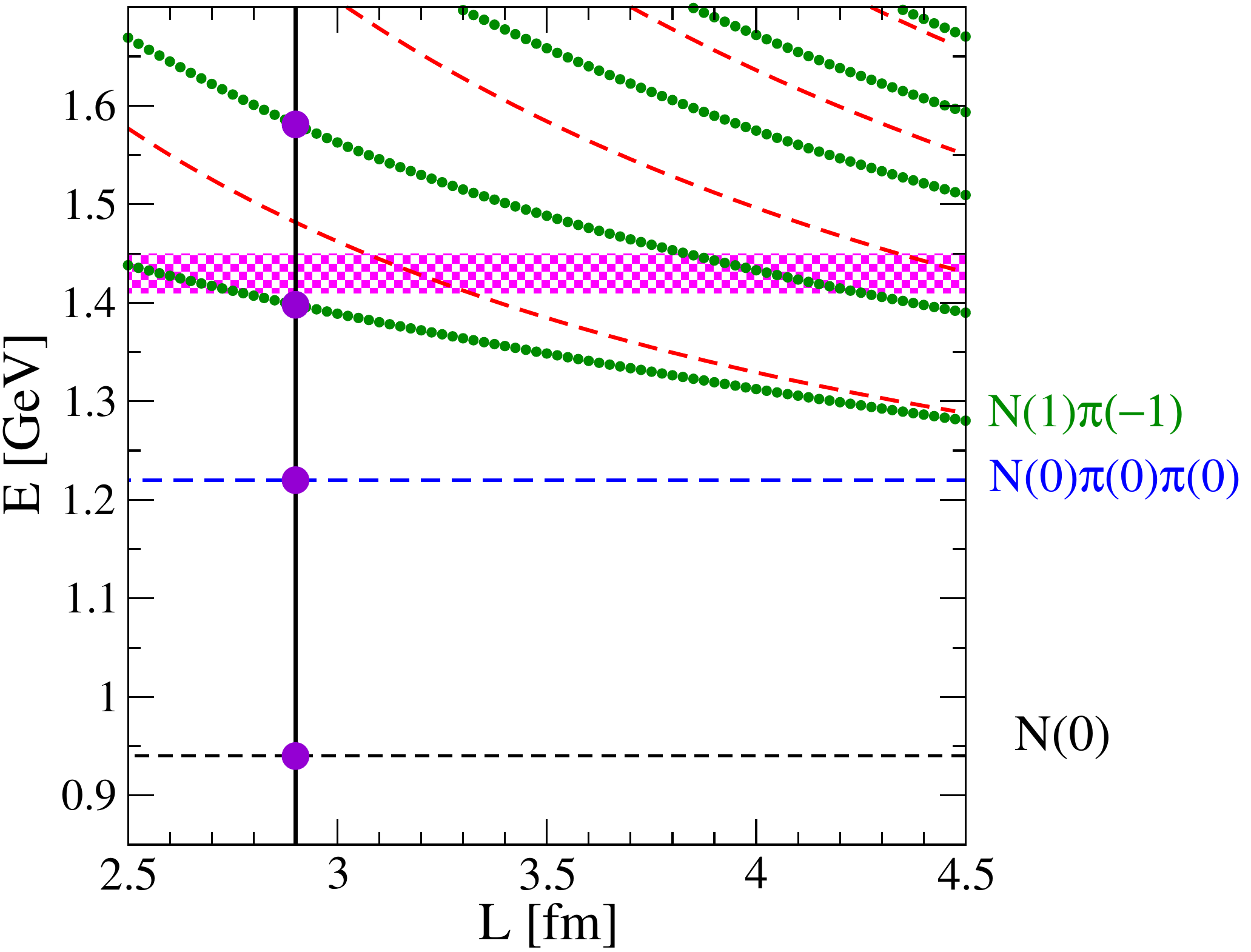}
\includegraphics[height=4.5cm,width=6.7cm]{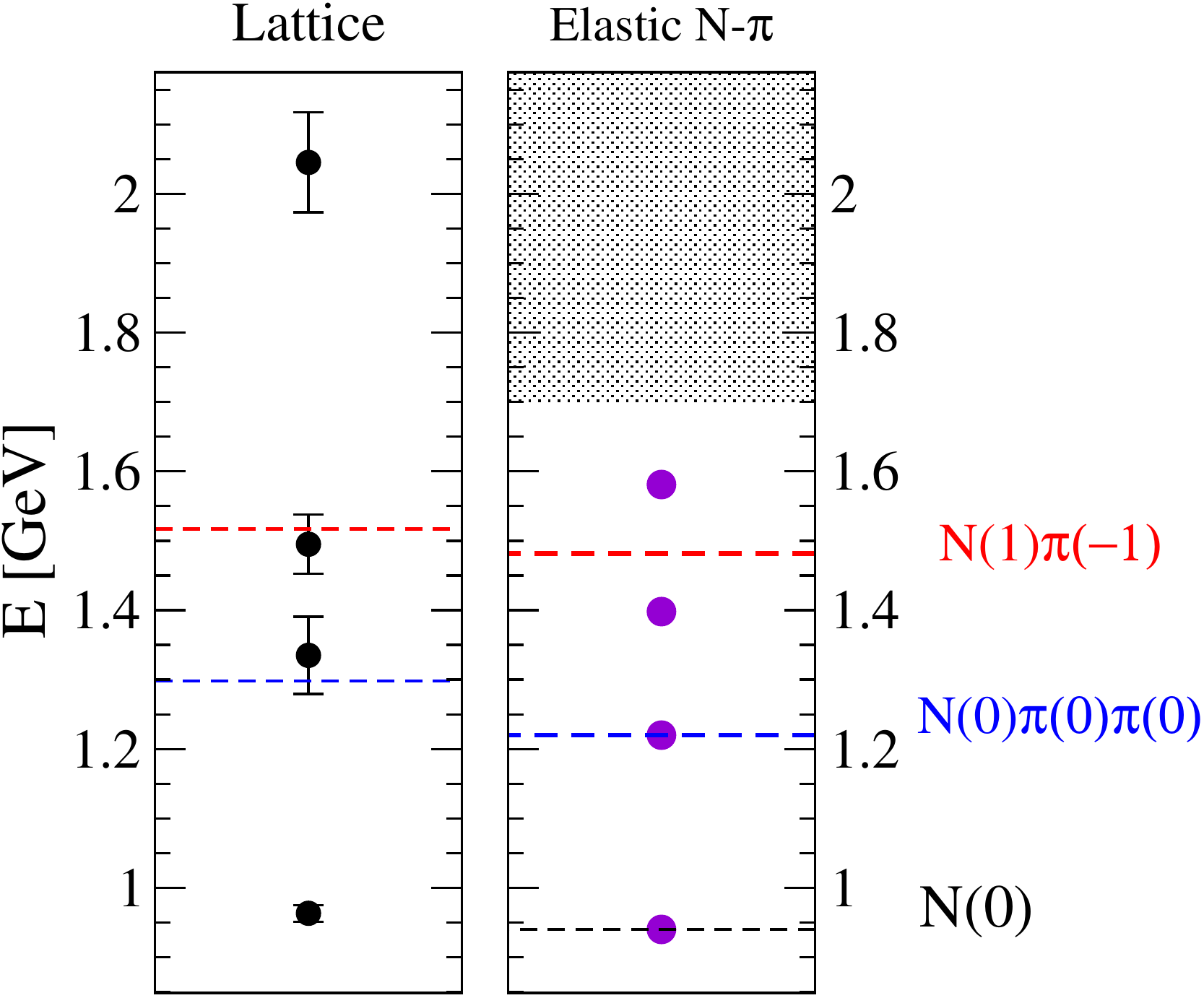}
\caption{Figures adapted from Ref. \cite{Lang:2016hnn}. Left: Excited energy spectrum of nucleons as a function of lattice size 
obtained by solving the inverse L\"uscher finite volume problem for $m_{\pi}=156$ MeV. Black, blue and red dashed lines indicate 
the non-interacting level positions for nucleon, $N\pi\pi$, and $N\pi$ levels, respectively. Dotted green lines indicate the 
$N\pi$ level positions levels in the presence of a Roper-like resonance coupled to them on the lattice. Right: Comparison of the 
excited nucleon spectrum between analytical expectations with numerical results from a lattice calculation with a pion mass 
156 MeV and physical spatial extension 2.9 fm. }\label{Npifig}
\end{figure}

In this regard, Ref. \cite{Lang:2016hnn} reports on an attempt to determine the excited spectrum of nucleon, including the scattering 
levels, and perform an elastic phase shift analysis within the Roper resonance energy region. This calculation studies only the spectrum 
in the rest frame of the nucleon and has been performed only on a single PACS-CS ensemble with $N_f=2+1$ dynamical flavors of Wilson clover 
fermions with $m_{\pi}=156$ MeV and physical lattice size $L=2.9$ fm. In a comparative study between the numerical results and theoretical 
expectation based on L\"uscher's finite volume method, they find the energy levels in the spectrum to be consistent with the expected 
non-interacting level positions (see Fig. \ref{Npifig}). This indicates the low lying Roper resonance does not arise on the lattice from 
the elastic $N\pi$ scattering. They also find signatures in operator state overlaps indicating strong coupled channel effects like those 
discussed in Ref. \cite{Kiratidis:2016hda,Wu:2017qve}, where the Roper resonance was described as a dynamically generated resonance due 
to coupled channel effects between $N\pi$, $N\sigma$ and $\Delta\pi$. This indeed calls for a more rigorous coupled channel investigation. 
Another interesting direction to pursue is to investigate the role of chiral symmetry in the excited nucleon spectrum, considering the fact 
that the only calculations using chiral fermions give low lying first excitations of the nucleon in the energy regime of the Roper resonance. 

\section{Beyond mesons and baryons}\label{beyond}

Many resonance structures, generally referred to as XYZs, have been discovered in the heavy quarkonium energy regime with 
properties contradicting the expectations from simple theoretical models. Starting with the discovery of $X(3872)$ in 
$B^{\pm}\rightarrow K^{\pm}X$ ($X\rightarrow J/\psi\pi^+\pi^-$) decays by Belle in 2003 \cite{Choi:2003ue}, currently there are 
several such candidates with an ambiguous nature in the charm and in the bottom sectors. Recently there has also been 
an observation of baryons in the $\Lambda_b^0\rightarrow J/\psi K^-p$ decays by LHCb \cite{Aaij:2015tga}, that are interpreted as 
charmonium-nucleon pentaquarks. A recent summary of efforts to find a theoretical description of these states can be found in 
Refs. \cite{Lebed:2016hpi,Esposito:2016noz,Olsen:2017bmm}.

{\bf Charmed tetraquarks}: Early lattice calculations assuming elastic $D\bar D$ scattering extracted a bound state pole and argued it to be the lattice 
candidate for $X(3872)$ \cite{Prelovsek:2013cra,Padmanath:2015era}. In these calculations, the lattice levels are associated
with non-interacting levels considering their nearness with the respective non-interacting level positions and the operator 
state overlaps. These studies followed a strategy of associating any additional energy level in the interacting spectrum, beyond 
those expected in the non-interacting spectrum, to indicate the presence of a narrow resonance. Alternatively, the deficiency 
of such additional levels is argued to indicate the absence of any resonance, {\it e.g.} in the case of hidden charm $I=1$ sector 
\cite{Prelovsek:2013xba}. Recently HSC has performed a detailed calculation to extract the finite volume spectra in the rest 
frame for $I=1$ hidden charm as well as doubly charm sectors using large bases of meson-meson and tetraquark interpolators 
\cite{Cheung:2017tnt}. Similar to other calculations referred to above, this calculation also followed the strategy of associating
the extracted lattice energy levels with the expected non-interacting meson-meson energy levels in the region considered. They also 
do not find any strong signatures for the presence of bound states or narrow resonances in the channels studied. All of these lattice 
calculations have been limited to zero momentum. It is argued in Ref. \cite{Padmanath:2015era} that the tetraquark interpolators 
utilized are related to the meson-meson interpolators via Fierz relations \cite{Nieves:2003in} and interpreting lattice levels 
based on their overlaps with tetraquark interpolators is subtle. In a recent letter by the HAL QCD Collaboration using their finite 
volume formalism, they investigate the interactions between $\pi J/\psi$, $\rho\eta_c$ and $\bar DD^{\ast}$ and argue that the charged 
$Z_c(3900)$ as a threshold cusp \cite{Ikeda:2016zwx}. 

\begin{figure}
\centering
\includegraphics[height=4.5cm,width=7.6cm]{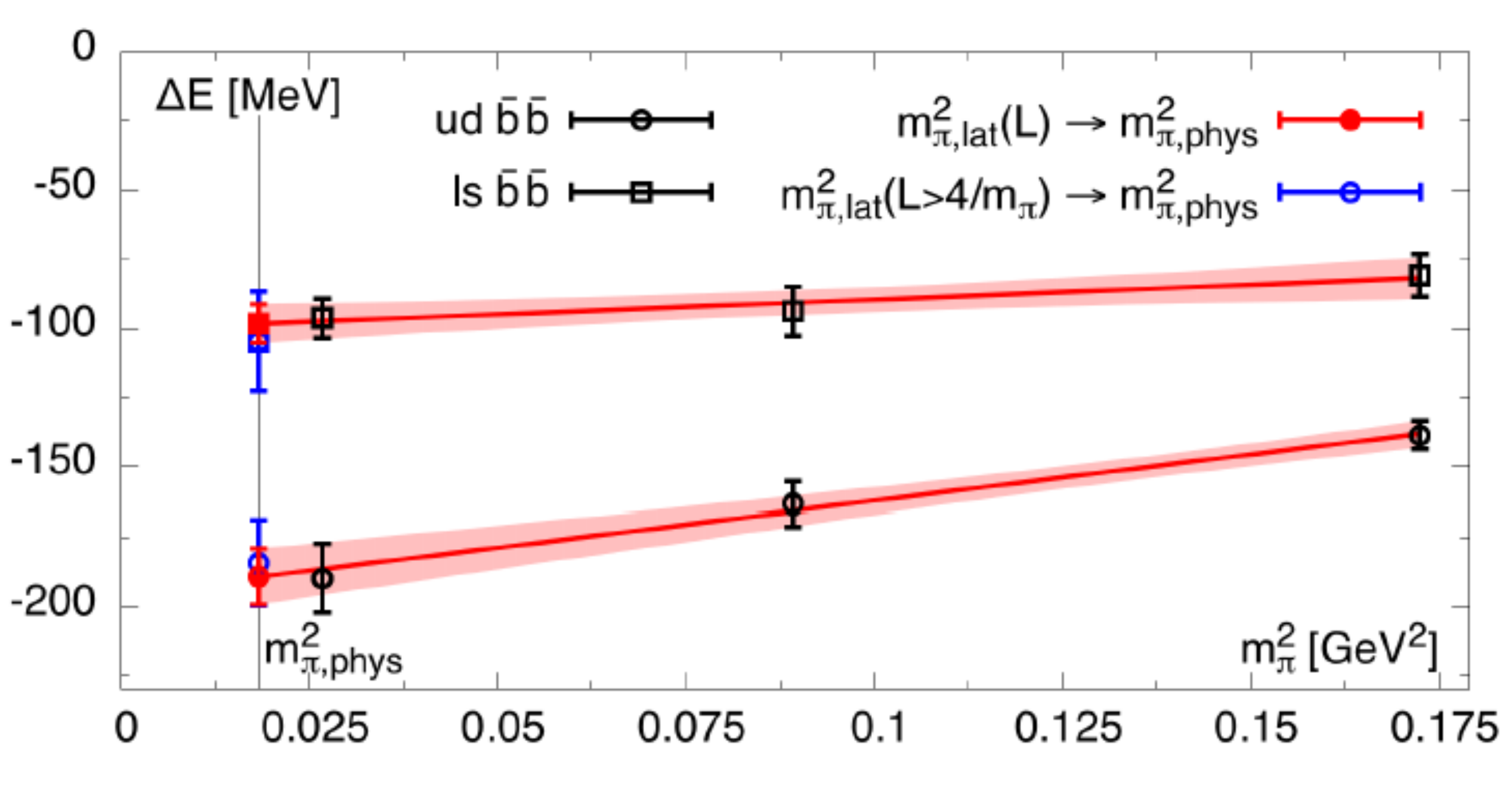}
\includegraphics[height=4.3cm,width=6.7cm]{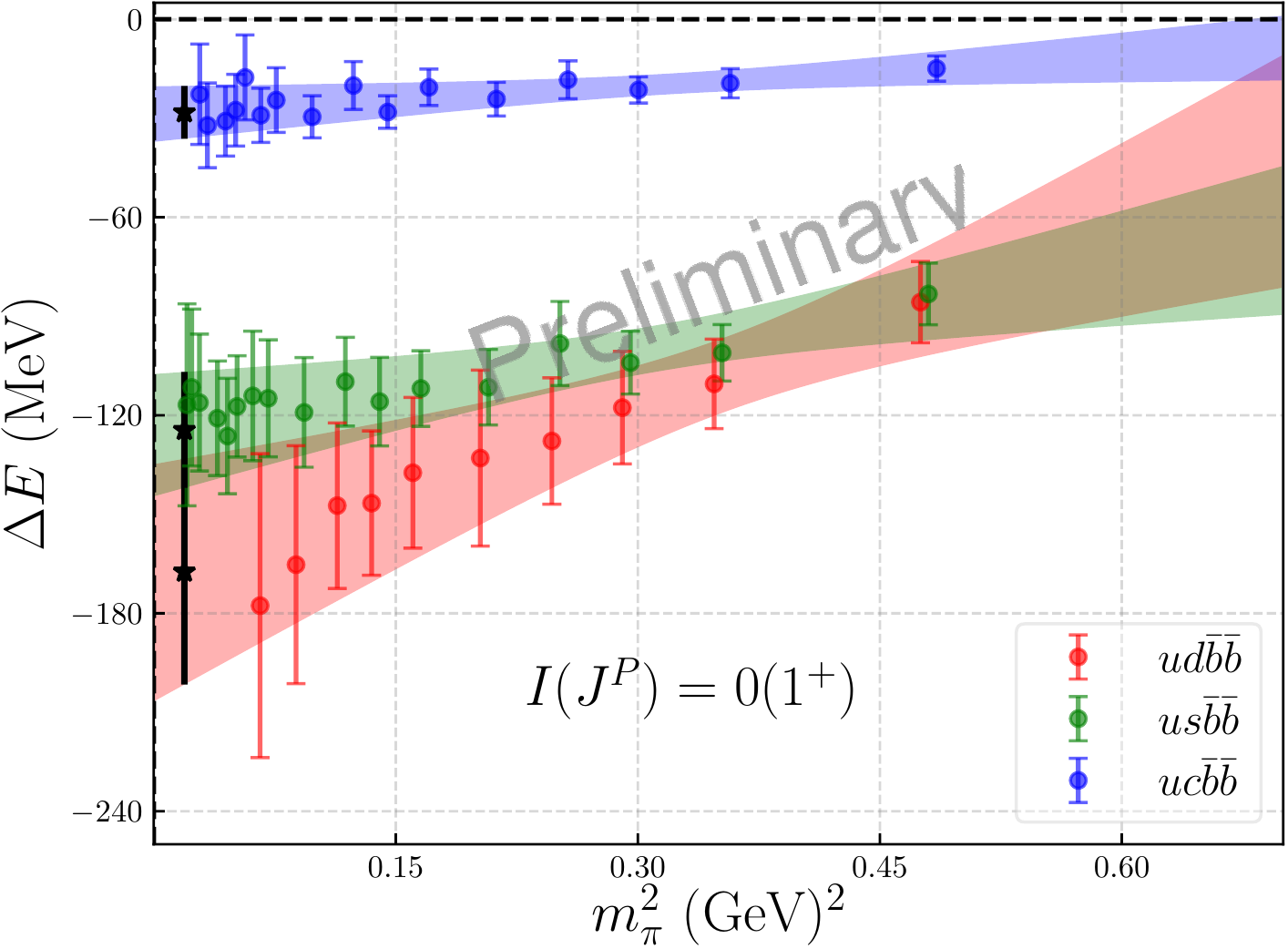}
\caption{Results for binding energies of doubly bottom axialvector tetraquarks with different flavor contents as a function of $m_{\pi}^2$ from 
Ref. \cite{Francis:2016hui} (left) and Ref. \cite{Junnarkar:2017sey} (right).}\label{bbtq}
\end{figure}

{\bf Doubly bottom tetraquarks}: The existence of stable doubly heavy tetraquark states has been proposed using potential model calculations \cite{Carlson:1987hh,
Karliner:2017qjm} and heavy quark symmetry \cite{Eichten:2017ffp}, for sufficiently large heavy quark mass. These calculations 
rely on the large mass of the heavy quark and so doubly bottom four-quark systems are perhaps more interesting than doubly 
charm systems. This has motivated many lattice groups to perform investigations of these systems \cite{Bicudo:2016ooe,Bicudo:2017szl,
Francis:2016hui}. Calculations in Ref. \cite{Bicudo:2016ooe} proceed by computing the potential of two static quarks in the 
presence of two light quarks, followed by solving the Schr\"odinger equation within Born-Oppenheimer approximation to study 
existing $\bar b\bar bqq$ states. In the $I(J^P)=0(1^+)$ channel, they find a bound state $\Delta E=90^{+43}_{-36}$ below the 
$BB^{\ast}$ threshold. An extension of this work reported the existence of a tetraquark resonance for $l=1$, decaying into two 
B mesons, with quantum numbers $I(J^P)=0(1^-)$ \cite{Bicudo:2017szl}. In Ref. \cite{Francis:2016hui}, the authors study the 
axial-vector channel for $I=1$ ($\bar b\bar bud$) and $I=1/2$ ($\bar b\bar bsu$) using non-relativistic QCD for the bottom quarks 
and find unambiguous signals for deeply bound doubly bottom tetraquarks with binding energies 189(10) MeV and 98(7) MeV, 
respectively (see the left of Fig. \ref{bbtq}). Recently the ILGTI Collaboration has also reported a similar calculation to 
investigate the light quark mass dependence of these binding energies \cite{Junnarkar:2017sey} arriving at similar conclusions. 
The plot in the right of Fig. \ref{bbtq} shows the binding energy as a function of $m_{\pi}$ for $J^P=1^+$ doubly bottom tetraquarks 
with different flavor contents. 

\begin{figure}
  \centering
  \incfig{0.54}{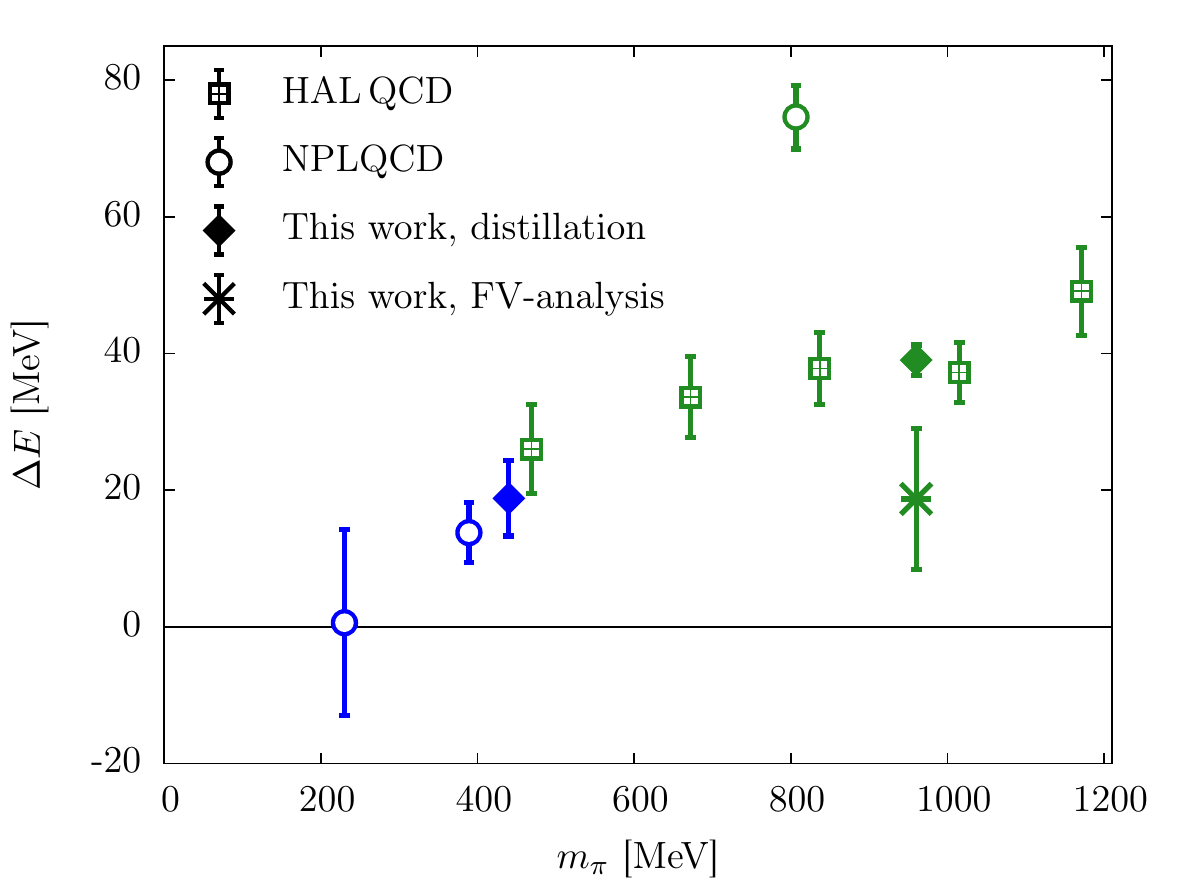}
\caption{(Figure adapted from Ref. \cite{Hanlon:2018yfv}) Comparison of the binding energy of $H$-dibaryon between different lattice calculations. 
Green and blue colors refer to calculations at the SU(3) symmetric point and the SU(3) broken cases.}\label{Hdibaryons}
\end{figure}

{\bf Penta quark systems}: Investigations by NPLQCD have provided interesting evidence for a shallow bound state in $\eta_c N$ system at the $SU(3)$ flavor 
symmetric point \cite{Beane:2014sda}. Recent studies of the effects of the light hadron cloud (with pion mass as low as $223$ MeV) 
on the potential between a static quark-antiquark pair have indicated many of these systems are energetically favorable with 
binding energies of less than a few MeV \cite{Alberti:2016dru}. Preliminary results for elastic $J/\psi-N$ scattering are 
reported at this meeting \cite{Skerbis:2018unn,Sugiura:2017vks}. Motivated by the recent discovery of two pentaquark candidates with 
spin 3/2 and 5/2 with opposite parities \cite{Aaij:2015tga}, studies are performed to extract the finite volume spectrum of the 
charmonium-nucleon system in the rest frame for both the parities \cite{Skerbis:2018unn}. In Ref. \cite{Sugiura:2017vks}, the authors
perform lattice QCD study of $J/\psi-N$ and $\eta_c-N$ systems using HAL QCD finite volume method and find these channels to be weakly
attractive, although the attraction is not strong enough to have a bound state. 

{\bf Six quark systems}: Several contributions at this meeting discuss recent lattice calculations exploring baryon-baryon interactions. 
The HAL QCD Collaboration reports on the recent updates of their investigations of $N\Omega$ interactions at near physical pion mass 
\cite{Iritani:2018sra}. In the $s$-wave spin 2 channel, from an $N_f=2+1$ flavor large volume lattice QCD simulation they find 
the possibility for a shallow quasi-bound state in the $N\Omega$ channel. In another recent article, HAL QCD report their study of 
$\Omega\Omega$ interactions resulting in shallow binding energies \cite{Gongyo:2017fjb}. In Ref. \cite{Francis:2018qch}, lattice 
group in Mainz report on their recent studies to resolve the question on whether $uuddss$ ($H$ dibaryon) system \cite{Jaffe:1976yi} 
is bound or not. There exist other lattice calculations performed in this regard by NPLQCD and HALQCD. All these early 
calculations reported a bound $H$ dibaryon at heavier than physical pion masses.  In Ref. \cite{Francis:2018qch}, the authors perform 
a $N_f=2$ flavor calculation using baryon-baryon interpolators as well as hexaquark operators. Performing a finite volume analysis, 
they find a bound $H$ dibaryon with a binding energy $\Delta E=19\pm10$ MeV for a pion mass of 960 MeV. Comparison 
of various lattice results is made in Fig. \ref{Hdibaryons}.

\section{Summary}\label{summar}

Precision measurements of ground state hadrons are now well established using lattice QCD. Several calculations are quite successful 
in precisely postdicting/predicting the masses of ground state hadrons, composed of $u$, $d$, $s$, $c$ and $b$ quarks, that are 
well below the strong decay threshold. A summary of lattice calculations of the ground state baryons can be found in Fig. \ref{lcbaryons} 
and Fig. \ref{hbaryons}. Notably, the mass of the recently discovered doubly charm baryon $\Xi_{cc}$ is in very good agreement with 
all the existing lattice predictions, demonstrating the ability of lattice QCD techniques to make reliable predictions. Exploratory 
calculations of excited baryon spectroscopy have been performed in a number of publications in the past 10 years. The masses and 
quantum numbers of the recently discovered tower of $\Omega_c$ resonances by the LHCb Collaboration have been successfully predicted 
by such lattice investigations.

Several lattice calculations have been reported of elastic scattering of spinless particles to study the simplest hadronic
resonances in the light, strange and charm meson sectors. Often most hadrons have several decay channels including scattering 
particles that have non-zero spin. Most of them are also open to 3-particle scattering channels, for which there is as yet no 
complete formalism that relates the three body scattering amplitudes with the discrete spectrum in a box. However, there is 
significant progress towards developing such a formalism. Calculations of coupled 2-particle scattering channels have been 
performed, where the pole singularities describing the scattering amplitudes are extracted by parameterizing the scattering 
matrix of these coupled channels. Recently there has also been a calculation, considering the scattering of mesons with 
non-zero spin. The first calculations involving finite volume analysis of the $\Delta$ baryon and the Roper resonance are also reported 
in this review.

Motivated by experimental evidence for the existence of tetraquarks in the hidden charm sector, the recent lattice calculations 
including the relevant meson-meson interpolators as well as tetraquark interpolators are presented. Investigations
of the ground state doubly bottom four-quark systems, inspired by phenomenological predictions, indicate deeply bound
tetraquark states in the axial-vector channels. Lattice studies of hadron interactions in the $J/\psi-N$ channel and dibaryon 
systems were also briefly reviewed.

Continuing lattice QCD efforts, to study the hadron spectrum on several volumes including all relevant multi-hadron interpolators 
and to perform a rigorous finite volume analysis, are necessary to further understand these resonances. Considering the recent 
progress in hadron spectroscopy using lattice QCD calculations, a time when we could understand the experimentally observed 
hadron resonance structures directly from Quantum ChromoDynamics is not very far.

\acknowledgments

I thank R. Brice\~no, J. Bulava, S. Cali', J. Dudek, G. Endrodi, K.F. Liu, A. T. Lytle,
A. Palasseri, S. Paul, G. Rend\'on, U. Skerbis, F. Stokes, A. Veerappan and H. Wittig for 
sharing information and valuable inputs to this review. I am grateful to G. S. Bali, S. Collins, 
R. G. Edwards, P. Junnarkar, C. B. Lang, L. Leskovec, N. Mathur, D. Mohler, S. Mondal, 
M. Peardon, S. Piemonte, S. Prelovsek, A. Sch\"afer and S. Weish\"apl for the pleasure in 
collaborating on various topics related to this review. In particular, I thank S. Collins and 
N. Mathur for various discussions in relation to this review. I acknowledge support from the 
EU under grant no. MSCA-IF-EF-ST-744659 (XQCDBaryons) and the Deutsche Forschungsgemeinschaft 
under grant No. SFB/TRR 55. I apologize for skipping some of the interesting calculations presented 
in this meeting due to time constraints. 

\bibliographystyle{JHEP}
\bibliography{references}

\end{document}